\def\DIRvalue{Pufu}
\def\IDvalue{PU}
\def\titlevalue{The $F$-Theorem and $F$-Maximization}
\def\authorvalue{Silviu S.~Pufu}
\def\shortauthorvalue{\authorvalue}
\def\addressvalue{Joseph Henry Laboratories, Princeton University, Princeton, NJ 08544, USA\\
  \tt spufu@Princeton.EDU}

\def\abstractvalue{This contribution contains a review of the role of
  the three-sphere free energy $F$ in recent developments related to the
  $F$-theorem and $F$-maximization.  The $F$-theorem states that for any
  Lorentz-invariant RG trajectory connecting a conformal field theory
  CFT${}_\text{UV}$ in the ultraviolet to a conformal field theory
  CFT${}_\text{IR}$, the $F$-coefficient decreases:  $F_\text{UV} >
  F_\text{IR}$.  I provide many examples of CFTs where one can compute
  $F$, approximately or exactly, and discuss various checks of the
  $F$-theorem.  $F$-maximization is the principle that in an ${\cal N} =
  2$ SCFT, viewed as the deep IR limit of an RG trajectory preserving
  ${\cal N} = 2$ supersymmetry, the superconformal R-symmetry maximizes
  $F$ within the set of all R-symmetries preserved by the RG trajectory.
  I review the derivation of this result and provide examples.
}

\def\preprintvalue{PUPT-2493}

\ifx\ifLONG\undefined
\documentclass[12pt]{article}

\newcommand{\chapterauthor}[1]{
\begin{center}
{\bf \normalsize  #1}
\end{center}
}

\newcommand{\chapteraddress}[1]{
\begin{center}
{ \small \it \addressvalue}
\end{center}
}

\newcommand{\chapterabstract}[1]{
\vspace{\baselineskip}
\begin{center}
\textbf{\small Abstract}
\end{center}
#1}

\newcommand{\chapterheader}{

\chapter[\titlevalue{}  (by \shortauthorvalue)]{\titlevalue}
\label{Chapter\IDvalue}
\chapterauthor{\authorvalue}
\chapteraddress{\addressvalue}
\chapterabstract{\abstractvalue}
\tightmtctrue
\minitoc
}

\newcommand{\documentheader}{
\begin{flushright} \small
  \preprintvalue
 \end{flushright}

\begin{center}
{\bf \Large \titlevalue}
\end{center}

\chapterauthor{\authorvalue}
\chapteraddress{\addressvalue}
\chapterabstract{\abstractvalue}

\medskip

This is a contribution to the review volume ``Localization techniques
in quantum field theories'' (eds. V.~Pestun and M.~Zabzine) which
contains 17 Chapters available at \cite{ContributionSummary}

\tableofcontents
}

\newcommand{\ifvolume}[2]{\ifx\ifLONG\undefined#2\else#1\fi}

\newcommand{\documentfinish}{
\ifx\ifLONG\undefined
\bibliographystyle{bibreview} 
\bibliography{\IDvalue,review}  
\end{document}
\else
\addcontentsline{toc}{section}{References}
\providecommand{\href}[2]{#2}\begingroup\raggedright\endgroup

\fi
}

\newcommand{\documentfinishBBL}{
\addcontentsline{toc}{section}{References}
\ifx\ifLONG\undefined
\input{\IDvalue.separate.bbl}
\end{document}
\else
\input{\DIRvalue/\IDvalue.volume.bbl}
\fi
}

\ifx\ifLONG\undefined
\def\volcite#1{Contribution \cite{Contribution#1}}
\else 
\def\volcite#1{Chapter \ref{Chapter#1}}
\fi

\usepackage{color}
\usepackage{amsmath}
\usepackage{amssymb}

\usepackage{graphicx}
\usepackage{caption}
\usepackage{epstopdf}
\usepackage{enumerate}
\usepackage{cite}
\usepackage{tensor}
\usepackage{slashed}

\usepackage{bbm} 
\usepackage{amsthm}
\usepackage{array}
\usepackage{mathtools}
\usepackage{dsfont}
\usepackage{multirow}
\usepackage{hyperref}
\usepackage{fullpage}

\numberwithin{equation}{section}

\begin{document}
\thispagestyle{empty}
\documentheader

\newcommand{\mop}[1]{\mathop{\rm #1}\nolimits}
\newcommand{\tr}{\mop{tr}}
\newcommand{\Tr}{\mop{Tr}}
\newcommand{\abs}[1]{\left\lvert #1 \right\rvert}
\newcommand{\BZ}{\mathbb{Z}}
\newcommand{\BR}{\mathbb{R}}
\newcommand{\BH}{\mathbb{H}}
\newcommand{\Det}{{\rm Det}}

\else 
\chapterheader 
\newcommand{\mop}[1]{\mathop{\rm #1}\nolimits}
\fi

\newcommand{\PUes}[2] {\begin{equation} \label{#1} \begin{split} #2 \end{split} \end{equation}}

\newcommand{\PUSSP}[1]{{}}
\newcommand{\Vol}{\mop{Vol}}

%ytableausetup{aligntableaux=center}

\numberwithin{equation}{section}

\section{Introduction}
\label{PUintro}

A central problem in theoretical physics is to find the laws that govern the low-energy dynamics of a physical system whose microscopic description is either fully known or, more often than not, only approximately known.  The conceptual framework in which this problem is usually formulated is that of the Renormalization Group (RG) flow---a motion in the space of theories that captures the effective degrees of freedom as they change with the energy scale used to probe the system.  Based on the intuition that the effective number of degrees of freedom decreases during this process, one expects that there exist quite general constraints that limit the set of possible RG trajectories.  In practice, such constraints could be extremely valuable because they could rule out or predict the kinds of low-energy dynamics that a given system might exhibit.   The goal of this Chapter is to discuss such a constraint, called the $F$-theorem, which applies to relativistic RG flows in three space-time dimensions, as well as to review a few related developments.

 If one insists on Lorentz invariance, as we will do in the present Chapter, a common picture is that the RG flow interpolates between two scale-invariant theories:  one valid at high energies (the ultraviolet fixed point) and one valid at low energies (the infrared fixed point).  In many cases, the symmetry of these scale-invariant theories is augmented to the conformal group, which in addition to Lorentz transformations and dilatations also includes special conformal transformations.\footnote{There are known exceptions, however. For instance, free Maxwell theory in three dimensions is scale invariant, but not conformally invariant in the ultraviolet.  See for example \cite{PUJackiw:2011vz,PUDymarsky:2013pqa, PUDymarsky:2014zja}.} Let us therefore restrict our attention to RG flows between two conformal field theories (CFTs).  The statement of the $F$-theorem is that each CFT can be assigned a number $F$, equal to the regularized $S^3$ free energy, $F \equiv -\log \abs{Z_{S^3}}$, as will be discussed in more detail shortly, with the property that whenever there exists an RG flow between a UV CFT and an IR CFT, one has $F_\text{UV} > F_\text{IR}$.  In other words, the $F$-coefficient always decreases under RG flow.   Consequently, the RG flow is not reversible:  if there exists a relativistic RG flow between CFT$_1$ in the UV and CFT$_2$ in the IR, a relativistic RG flow between CFT$_2$ in the UV and CFT$_1$ in the IR is ruled out.

The $F$-theorem is a rather recent development that applies to three-dimensional QFTs.  It has older analogs in other numbers of space-time dimensions.  For instance, in two space-time dimensions, Zamolodchikov \cite{PUZamolodchikov:1986gt} showed that the central charge $c$ always decreases under RG flow, a result known as the $c$-theorem.  Similarly, in four-dimensional theories, Cardy \cite{PUCardy:1988cwa} conjectured that the Weyl anomaly coefficient $a$ has a similar property.  This result is known as the $a$-theorem and was proven recently by Komargodski and Schwimmer \cite{PUKomargodski:2011vj}.   Likewise, in one space-time dimension (i.e.~for quantum mechanical systems) it has been proven that the thermal free energy decreases under RG flow, a statement known as the $g$-theorem \cite{PUAffleck:1992ng,PUAffleck:1991tk}.  

There is a significant difference between even and odd space-time dimensions, however.  The coefficients $c$ and $a$ in two and four space-time dimensions, respectively, correspond to anomalies that are not present in odd space-time dimensions.  These anomaly coefficients appear in correlation functions at separated points and can therefore be calculated from correlation functions of local operators.  The $F$-coefficient, on the other hand, is a non-local quantity.  Indeed, it can be non-vanishing even in topological field theories that have no local degrees of freedom, as will be shown in the case of Chern-Simons theory in Section~\ref{PUCHERNSIMONS}.   Despite these differences between even and odd space-time dimensions, a unified treatment in all dimensions can be achieved by considering the sphere partition function of the Euclidean CFT\@.

There are two different paths that led to the development of the $F$-theorem.  The first path involves supersymmetric gauge theories.  Three-dimensional superconformal field theories (SCFTs) with ${\cal N} = 2$ supersymmetry possess a global $U(1)_R$ R-symmetry\footnote{An R-symmetry is a symmetry that does not commute with the supersymmetry generators.} that appears in the superconformal algebra, just as in the case with ${\cal N} = 1$ SCFTs in four dimensions.  This symmetry is important because it determines various unitarity bounds---for instance, the scaling dimensions of all scalar operators must be at least equal to the magnitude of their $U(1)_R$ charge, and scalar operators that saturate this bound have very special properties.  Given an SCFT, it is desirable to determine the $U(1)_R$ charges of the various operators of the theory.  

In general, the determination of the $U(1)_R$ symmetry is a very difficult problem even for theories with Lagrangian descriptions where the SCFT of interest is realized as the infrared limit of an RG flow, because generically the $U(1)_R$ R-symmetry appears as an emergent symmetry only at the IR fixed point.  In cases where the SCFT of interest can be reached via an ${\cal N} = 2$-preserving RG flow that does preserve some R-symmetry $U(1)_R^\text{RG}$ and there are no accidental symmetries at the IR fixed point, the $U(1)_R$ symmetry that appears in the superconformal algebra of the IR SCFT can be a linear combination of $U(1)_R^\text{RG}$ and other abelian flavor symmetries preserved throughout the RG flow.  It was proposed by Jafferis in \cite{PUJafferis:2010un} and later refined in \cite{PUJafferis:2011zi,PUClosset:2012vg} that a procedure called $F$-maximization can be used to determine this linear combination.  As will be discussed in more detail in Section~\ref{PUFMAXIMIZATION}, the procedure involves maximizing the $S^3$ free energy $F$ over a family of QFTs.  A similar procedure for ${\cal N} = 1$ theories in four dimensions called $a$-maximization \cite{PUIntriligator:2003jj}  had been known previously and involved maximizing the anomaly coefficient $a$.  Since the four-dimensional anomaly coefficient $a$ and the three-dimensional $S^3$ free energy $F$ play similar roles, and since the 4-d $a$-coefficient was expected to be monotonic under RG flow, it was proposed in \cite{PUJafferis:2011zi} that the $F$-coefficient should have the same property.  Many tests of this proposal were performed both in supersymmetric and non-supersymmetric RG trajectories \cite{PUJafferis:2011zi, PUGulotta:2011si, PUKlebanov:2011gs, PUKlebanov:2011td}.

The second path to the development of the $F$-theorem starts from the work of Myers and Sinha \cite{PUMyers:2010xs, PUMyers:2010tj}, who studied RG flows for quantum field theories with holographic duals.  For Lorentz-invariant holographic RG flows between $d$-dimensional CFTs, they identified a quantity $a_d$ that decreases monotonically.  Let $a_d^*$ denote the value of $a_d$ at an RG fixed point.  It can be shown that in $d=2$ one has $a_2^* = c$ and in $d=4$ one has $a_4^* = a$, and thus the quantity $a_d^*$ provides a generalization of these conformal anomaly coefficients to arbitrary $d$.  As explained in \cite{PUMyers:2010tj}, the quantity $a_d^*$ can also be interpreted as the universal part in the vacuum entanglement entropy between a ball of radius $R$ and its complement.  (In general, this entanglement entropy is divergent, being proportional to the area of the boundary of the ball in units of the UV cutoff.  In a CFT it is possible to subtract unambiguously the power divergences and identify a universal contribution.)    The connection between the three-dimensional $a_3^*$ and the $F$-coefficient as defined above was made in \cite{PUCasini:2011kv}, where it was shown that for any CFT, the universal part of the vacuum entanglement entropy between a ball and its complement is equal precisely to the $F$-coefficient.

This second line of development was taken further by Liu and Mezei \cite{PULiu:2012eea}, who defined for any 3-d Lorentz-invariant RG flow the notion of renormalized entanglement entropy (REE), which is a finite quantity which agrees with the $F$-coefficient at the UV and IR fixed points.  As shown further by Casini and Huerta \cite{PUCasini:2012ei}, the REE interpolates monotonically between $F_\text{UV}$ and $F_\text{IR}$ in any Lorentz-invariant RG flow, thus establishing the $F$-theorem.  This proof of the $F$-theorem is based on the strong subadittivity property of entanglement entropy and generalizes a similar proof of the $c$-theorem in 2-d \cite{PUCasini:2006es}.   Extending this proof to dimensions higher than three is currently still an open problem. 

In the rest of this Chapter I provide more details on these recent developments.  I focus on the definition of $F$ as the universal part in the $S^3$ free energy as opposed to its definition in terms of entanglement entropy, partly because this definition is more closely related to the topics presented in this volume, and partly because it renders $F$ more easily computable in many cases.  In Section~\ref{PUFCOEFFICIENTS} I discuss the computation of the $F$-coefficient in free theories and in various approximations such as the $1/N$ expansion or the $\epsilon$ expansion, as well as in supersymmetric theories.  These computations provide many checks of the $F$-theorem.  In Section~\ref{PUFMAXIMIZATION} I discuss the $F$-maximization principle and show how it can be applied in a few simple examples.

\section{The $F$-coefficients of various CFTs}
\label{PUFCOEFFICIENTS}

\subsection{$F$ in free field theories}

Let us begin with presenting the computation of $F$ in free field theories, where $F$ can be easily calculated by performing a Gaussian integral \cite{PUQuineChoi,PUKumagai,PUKlebanov:2011gs}. 

\subsubsection{Free real scalar field}
\label{PUSCALAR}

To calculate the $F$-coefficient of a free massless real scalar field, we start with the action for a conformally coupled scalar:
 \PUes{PUActionScalar}{
  {\cal S}_S = \frac 12 \int d^3 x \sqrt{g} \left[ \partial_\mu \phi \partial^\mu \phi + \frac {\cal R}8 \phi^2 \right] \,.
 }
The conformal coupling term, ${\cal R} \phi^2/8$, guarantees that this action is invariant under Weyl rescalings $g_{\mu\nu}(x) \to e^{2 \Omega(x)} g_{\mu\nu}(x)$, $\phi(x) \to e^{-\Omega(x)/2} \phi(x)$, for any $\Omega(x)$. Equivalently, the conformal coupling guarantees that the correlation functions of $\phi$ computed on a conformally flat space from the action \eqref{PUActionScalar} agree with those obtained by conformal transformation of the flat space ones.  On an $S^3$ of unit radius, we have that the Ricci scalar is ${\cal R} = 6$.  The $S^3$ free energy is therefore
 \PUes{PUFFreeScalar}{
  F_S = - \log \abs{Z_S} = \frac 12 \tr \log \left( -\nabla^2 + \frac 34 \right) \,.
 }
The operator $-\nabla^2 + \frac 34$ has eigenvalues 
 \PUes{Evalues}{
  n(n+2) + \frac 34 = \left(n+ \frac 12 \right) \left(n + \frac 32 \right) \,, \qquad n = 0, 1, 2, \ldots \,,
 }
each appearing with degeneracy $(n+1)^2$.  The free energy is thus
 \PUes{PUFFreeScalarAgain}{
  F_S = \frac 12 \sum_{n=0}^\infty (n+1)^2 \log \left[ \left(n+ \frac 12 \right) \left(n + \frac 32 \right) \right]  \,.
 }
This sum is cubically divergent, but it can be regularized in zeta-function regularization, or by simply multiplying each term by an exponential damping factor $e^{-\epsilon  \left(n+ \frac 12 \right) \left(n + \frac 32 \right)}$, with $\epsilon > 0$, and removing the power divergences in $1/\epsilon$ in an expansion at small $\epsilon$.  The regularized value of $F_S$ is \cite{PUQuineChoi, PUKumagai, PUMarino:2011nm}
 \PUes{PUGotFScalar}{
  F_S = \frac{\log 2}{8} - \frac{3 \zeta(3)}{16 \pi^2}  \approx 0.0638 \,.
 }

A notable feature of this answer is that it is strictly positive, as can be inferred from the $F$-theorem as follows.  While the theory of a free massless scalar considered above is a CFT, the theory of a free massive scalar is not.  Instead, a free massive scalar can be thought of as an RG trajectory connecting the CFT of a free massless scalar in the UV and an empty theory (a theory with no local or non-local operators) in the IR\@.  The empty theory is a trivial CFT that should be assigned $F_\text{empty} = 0$.  That $F_S > F_\text{empty} = 0$ is thus a necessary condition for the $F$-theorem to hold.

\subsubsection{Free Dirac fermion}
\label{PUDIRAC}

The next example of a free theory that we discuss is a free massless Dirac fermion (two-component complex spinor), for which the action on $S^3$ can be written as
 \PUes{PUActionFermion}{
  {\cal S}_D = \int d^3x \sqrt{g} \psi^\dagger (i \slashed{D}) \psi \,,
 }
where $\slashed{D}$ is the Dirac operator.  The free energy is
 \PUes{PUFFreeFermion}{
  F_D = -\log \abs{Z_D} = -\tr \log (i \slashed{D}) \,.
 }
As in the case of a massless scalar, writing $F_D$ as a sum over the eigenvalues of $i \slashed{D}$ yields a divergent answer that can be regularized, for instance, using zeta function regularization.  For a free massless Dirac fermion, the regularized value of $F_D$ is
 \PUes{PUGotFDirac}{
  F_D = \frac{\log 2}{4} + \frac{3 \zeta(3)}{8 \pi^2} \approx 0.219 \,.
 }
As in the scalar case discussed above, there exists an RG trajectory corresponding to a free massive fermion that connects the CFT of a free Dirac fermion to the empty theory for which $F_\text{empty} = 0$.  That $F_D > 0$ is consistent with the $F$-theorem.

For a free Majorana fermion, we have
 \PUes{PUGotFMajorana}{
  F_M = \frac 12 F_D = \frac{\log 2}{8} + \frac{3 \zeta(3)}{16 \pi^2} \approx 0.109 \,.
 }
A free minimal ${\cal N} = 1$ multiplet consisting of a free real scalar and a free Majorana fermion therefore has
 \PUes{PUGotFMinimal}{
   F_S + F_M = \frac{\log 2}{4} \approx 0.173 \,.
 }
The $F$-coefficients of a free massless ${\cal N} = 2$ chiral multiplet (consisting of a free massless complex scalar and a free massless Dirac fermion) and a free massless hypermultiplet (consisting of two massless complex scalars and two massless Dirac fermions) are twice and four times the value quoted in \eqref{PUGotFMinimal}, respectively:
 \PUes{PUGotFN2}{
  F_\text{chiral} = 2(F_S + F_M) = \frac{\log 2}{2} \approx 0.347 \,, \qquad
  F_\text{hyper} = 4(F_S + F_M) = \log 2 \approx 0.693 \,.
 }

\subsubsection{Chern-Simons theory}
\label{PUCHERNSIMONS}

Another case in which the $F$-coefficient can be computed is that of Chern-Simons theory at level $k$, described by the 3-d action
 \PUes{PUActionCS}{
  {\cal S}_\text{CS} = \frac{i k}{4 \pi} \int  \tr \left[ A \wedge dA - \frac {2i}3 A \wedge A \wedge A \right] \,,
 }
where the gauge field $A$ transforms in the adjoint representation of the gauge group.  In the case of $U(N)$ (or $SU(N)$) the Chern-Simons level $k$ is quantized in integer units provided that ``$\tr$'' is the trace in the fundamental representation.  In the $N=1$ case, an explicit evaluation of the Gaussian integral as in the scalar and fermion cases above, one obtains
In this case
 \PUes{PUGotFCS}{
  F_\text{CS} = \frac 12 \log k \,.
 }
More generally, for $U(N)$ Chern-Simons theory at level $k$, we have, when $N>1$, \cite{PUWitten:1988hf}
 \PUes{PUGotFUNCS}{
  F_\text{CS}(k, N) = \frac{N}{2} \log (k + N) - \sum_{j=1}^{N-1} (N-j) \log \left(2 \sin \frac{\pi j}{k + N} \right) \,.
 }

It is worth noting that in Maxwell theory on an $S^3$ of radius $R$, we have \cite{PUKlebanov:2011td} (see also \cite{PUAgon:2013iva})
 \PUes{PUFMaxwell}{
  F_\text{Maxwell} = - \frac{\log e^2 R}{2} + \frac{\zeta(3)}{4 \pi^2} \,.
 }
This theory is not conformal, because $e^2$ has dimensions of mass.  When $R \to 0$, we have $F_\text{Maxwell} \to \infty$, which is again in agreement with the $F$-theorem as applied to the case of the Maxwell-Chern-Simons theory, which can be thought of as an RG flow between Maxwell theory in the UV and Chern-Simons theory at level $k$ in the IR\@.

\subsection{$F$ in perturbative expansions}
\label{PUPERTURBATIVE}

In theories that are not free, the $F$-coefficient is more difficult to calculate.  As we now describe, it can be calculated approximately in various perturbative expansions.  In Sections~\ref{PUSUPERSYMMETRIC} and~\ref{PUFMAXIMIZATION}, we will describe the computation of $F$ in supersymmetric theories, where it can be computed exactly.

\subsubsection{Theories with many flavors}
\label{PULARGEN}

Theories with many flavors, such as the critical $O(N)$ vector model, become approximately free when the number of flavors is taken to infinity, and the $F$-coefficient of such a theory approaches that of the corresponding free theory.  A general class of such theories are Chern-Simons theories coupled to charged bosons and fermions, such as a $U(1)$ gauge theory with Chern-Simons level $k$ to which we couple $N_b$ complex scalars with charge $q_b$ and $N_f$ Dirac fermions with charge $q_f$ with the action on $\BR^3$ given by
 \PUes{PULagCSMatter}{
  {\cal S} &= \frac{i k}{4 \pi} \int A \wedge dA + \int d^3x\, \Biggl[
    \sum_{a=1}^{N_b} \left[ \abs{(\partial_\mu  - i q_b A_\mu) \phi_a}^2  + s \abs{\phi_a}^2 \right] \\
    &\qquad\qquad\qquad\qquad\qquad{}+ \sum_{\alpha=1}^{N_f} \psi^\dagger_\alpha (i \slashed{D} + q_f \slashed{A}) \psi_\alpha 
    + \frac{u}{2}  \left( \sum_{a=1}^{N_b} \abs{\phi_a}^2 \right)^2 \Biggr] + \ldots \,.
 }
Here, the ellipses denote other possible interactions, such as a Maxwell term for the gauge field, or a sextic boson coupling, which are believed to be irrelevant at large distances.  This action does not include fermion mass terms or Yukawa terms, as these can be projected out by imposing discrete symmetries.
 
An interesting limit of \eqref{PULagCSMatter} is that where $N_f$, $N_b$, and $k$ are all taken to infinity together.  In this limit, one can argue that, upon tuning the boson mass parameter $s$, the IR physics is described by an interacting conformal field theory.  To leading order, the $F$-coefficient is given by that of $N_b$ free complex scalars and $N_f$ free Dirac fermions, namely
 \PUes{PUFCSMatterLeading}{
  F = 2 N_b F_S + N_f F_D + \ldots \,, 
 }
where $F_S$ and $F_D$ are given in \eqref{PUGotFScalar} and \eqref{PUGotFDirac}, respectively.

The first correction to \eqref{PUFCSMatterLeading} can be computed as follows \cite{PUKlebanov:2011td}.  The first step is to decouple the quartic scalar interaction with the help of a dynamical Hubbard-Stratanovich field $\lambda$, thus making the replacement 
 \PUes{PUQuarticDecoupling}{
   \frac{u}{2}  \left( \sum_{a=1}^{N_b} \abs{\phi_a}^2 \right)^2 \to 
    i \lambda \sum_{a=1}^{N_b} \abs{\phi_a}^2 + \frac{\lambda^2}{2u} 
 }
in \eqref{PULagCSMatter}.  Integrating over $\lambda$, one recovers the quartic scalar interaction.  At low energies, $u$ grows and the term $\lambda^2/u$ can be dropped.  One can then consider the resulting action on any conformally flat space:
 \PUes{PUCSMatterS3}{
  {\cal S} &= \int d^3x\, \sqrt{g} \Biggl[
    \sum_{a=1}^{N_b} \left[ \abs{(\partial_\mu  - i q_b A_\mu) \phi_a}^2  + \left( \frac{\cal R}{8} + i \lambda \right)  \abs{\phi_a}^2 \right] \\
    &+ \sum_{\alpha=1}^{N_f} \psi^\dagger_\alpha (i \slashed{D} + q_f \slashed{A}) \psi_\alpha + \frac{i k}{4 \pi} \int A \wedge dA  \,.
 }
In the case of an $S^3$ of unit radius, we have ${\cal R} = 6$, as before.

Performing the Gaussian path integral over the bosons and fermions, the $S^3$ partition function becomes
 \PUes{PUS3Part}{
  Z  = \int DA_\mu\,  D\lambda\, \exp\left[ - {\cal S}_\text{eff}[A_\mu, \lambda] \right]  \,,
 }
with 
 \PUes{PUSeff}{
  {\cal S}_\text{eff}[A_\mu, \lambda] = N_b \tr \log \left(- (\nabla_\mu - i q_b A_\mu)^2 + \lambda + \frac 34 \right) 
   - N_f  \tr \log (i \slashed{D} + q_f \slashed{A}) + \frac{i k}{4 \pi} \int A \wedge dA \,.
 }
When $N_f$, $N_b$, and $k$ are large, the typical fluctuations of both the gauge field and the Lagrange multiplier are suppressed, and one can calculate $Z$ in a saddle point approximation:
 \PUes{PUZS3Saddle}{
  Z &\approx e^{-{\cal S}_\text{eff}[0, 0] } \int DA_\mu\,  D\lambda\, \exp\Biggl[ - \frac 12 \int d^3x d^3y A_\mu(x) A_\nu(y)
  \left( -\frac{\delta^2 {\cal S}_\text{eff}[A_\mu, \lambda]}{\delta A_\mu(x) \delta A_\nu(y)}\Biggr|_{A = \lambda =0} \right) \\
   &- \frac 12 \int d^3x d^3y \lambda(x) \lambda(y)
  \left( \frac{\delta^2 {\cal S}_\text{eff}[A_\mu, \lambda]}{\delta \lambda(x) \delta \lambda (y)}\Biggr|_{A = \lambda =0} \right) 
    \Biggr] \,,
 }
obtained by expanding the effective action \eqref{PUSeff} to quadratic order in the fluctuations and re-exponentiating.  (The first derivatives with respect to $\lambda$ and $A_\mu$, as well as the mixed second derivative vanish due to conformal invariance of the $A = \lambda = 0$ theory.)

The quantity ${\cal S}_\text{eff}[0, 0]$ reduces to the functional determinants in the free theory computed in the previous section.  It gives precisely
 \PUes{PUS00}{
  {\cal S}_\text{eff}[0, 0] = 2N_b F_S + N_f F_D \,.
 }
The Gaussian integral in \eqref{PUZS3Saddle} over $A_\mu$ and $\lambda$ is rather complicated, but was performed in \cite{PUKlebanov:2011td}.  The final result for $F$ is
 \PUes{PUFComplicated}{
  F = 2N_b F_S + N_f F_D + \frac 12 \log \left[ \pi \sqrt{\left(\frac{q_f^2 N_f + q_b^2 N_b}{8} \right)^2 + \left( \frac{k}{\pi} \right)^2 } \right] - \frac{\zeta(3)}{8 \pi^2} + \ldots \,,
 }
where the corrections are suppressed at large $N_f$, $N_b$, and $k$.

A particular case of \eqref{PUFComplicated} gives an approximate expression for the $F$-coefficient of the critical $O(N)$ vector model.  Indeed, setting $N_b = N/2$, as the fields appearing in the critical $O(N)$ model are real scalars, as well as setting $N_f =0$, $q_b = 0$, and removing the topological contribution $\frac 12 \log k$ from \eqref{PUFComplicated}, one obtains
 \PUes{PUFON}{
  F_\text{critical}(N) = N F_S - \frac{\zeta(3)}{8 \pi^2}  + O(1/N) \,.
 }

The correction to the free theory result given by the second term in \eqref{PUFComplicated} is a particular case of a more general setup, where instead of gauging a spin-$1$ current we gauge one of the higher spin currents of the free theory \cite{PUGiombi:2013yva}.  A related approximation scheme can be developed in the case of double trace deformations of large $N$ vector models \cite{PUKlebanov:2011gs,PUGiombi:2013yva}.

\subsubsection{$F$ from the $\epsilon$ expansion}
\label{PUEPSILON}

Another perturbative expansion that is commonly used in the theory of critical phenomena is the $\epsilon$ expansion \cite{PUWilson:1971dc} (for a review, see \cite{PUWilson:1973jj}).  This method is successfully used, for instance, to estimate certain critical exponents in CFTs such as the 3-d Ising model, and these estimates agree quite well with experimental measurements and lattice simulations.  The idea of the $\epsilon$ expansion is to continue the CFT of interest to non-integer space-time dimensions and identify the number of spacetime dimensions $d=d_*$ where the theory can becomes free, or more generally, where the theory can be solved exactly.  Then one can develop a perturbative expansion in $\epsilon = d - d_*$ and extrapolate these results to the value of $d$ of interest.

In recent work, Refs.~\cite{PUGiombi:2014xxa,PUFei:2015oha,PUGiombi:2015haa} applied the $\epsilon$ expansion method to the computation of the $F$-coefficient in theories such as the critical $O(N)$ vector models, Gross-Neveu models, and quantum electrodynamics with charged fermionic matter.  In general, it is expected that the free energy $F^{(d)}$ on a round $S^d$ should diverge (even after removing non-universal divergences) when $d$ approaches an even integer due to the presence of a conformal anomaly.  Indeed, when $d$ is an even integer, the $S^d$ free energy has a universal logarithmic dependence on the radius of the sphere proportional to the conformal anomaly coefficient---in $d=2$ the coefficient of the logarithmic divergence equals $\frac{\pi}{6} c$ in the convention where a free non-compact real scalar has $c=1$, while in $d=4$ this coefficient equals $\frac{\pi}{2} a$ in the convention where a free real scalar has $a = 1/90$.  The theories studied in \cite{PUGiombi:2014xxa,PUFei:2015oha,PUGiombi:2015haa} are solvable precisely in even $d$, where these divergences occur.  To obtain a finite quantity, Refs.~\cite{PUGiombi:2014xxa,PUFei:2015oha,PUGiombi:2015haa} defined
 \PUes{PUFTildeDef}{
   \tilde F^{(d)} = -F^{(d)} \sin \frac{\pi d}{2} \,.
 }
After removing non-universal divergences, the quantity $\tilde F^{(d)}$ is finite in all $d$.  It interpolates between $F$ in $d=3$ and the anomaly coefficients in $d=2$ and $d=4$.

The critical $O(N)$ vector model merges with the free CFT of $N$ scalar fields in $d=4$, and in $d=4-\epsilon$ dimensions, with $\epsilon \ll 1$, it is thus weakly coupled.  In~\cite{PUGiombi:2014xxa,PUFei:2015oha} it was found that 
 \PUes{PUFtON}{
  \tilde F_{O(N)}^{(4-\epsilon)} &= N \tilde F_S^{(4-\epsilon)} - \frac{\pi N (N+2)}{576 (N+8)^2} \epsilon^3 - \frac{\pi N (N+2)(13 N^2 + 370N + 1588}{6912 (N+8)^4} \epsilon^4 \\
    {}&+ \frac{\pi N (N+2)}{414720 (N+8)^6} 
     \biggl[  
       10368 (N+8)(5N+22)\zeta(3) - 647 N^4 - 32152 N^3\\
    {}&- 606576 N^2 - 3939520 N + 30 \pi^2 (N+8)^4 - 8451008
     \biggr] \epsilon^5 + O(\epsilon^6) \,,
 }
where 
 \PUes{PUFSDef}{
  \tilde F_S^{(d)} = \frac{1}{\Gamma(1 + d)} \int_0^1 du\, u \sin (\pi u) \, \Gamma \left( \frac d2 + u \right) \Gamma \left( \frac d2 - u \right) 
 }
represents the value of $\tilde F^{(d)}$ of a free conformally coupled scalar.  In $d=3$, \eqref{PUFSDef} reduces to \eqref{PUGotFScalar}.

Of particular interest is the $N=1$ case where the critical $O(N)$ vector model is nothing but the critical Ising theory.  In this case $O(1) = \BZ_2$, and the critical Ising theory can be thought of the IR fixed point of a real scalar with a quartic potential,
 \PUes{PULagIsing}{
  S_\text{Isign} = \int d^d x \, \left[(\partial_\mu \phi)^2 + m^2 \phi^2 + \lambda \phi^4 \right] \,,
 }
upon tuning the mass $m$ to zero.  When continued in $d$, it is well-known that the critical Ising theory has $c=1/2$ in $d=2$, corresponding to $\tilde F^{(2)} = \pi/12$ exactly.  Using this exact value in $d=2$ and \eqref{PUFtON} close to $d=4$, one can then construct an interpolating function and estimate that, in $d=3$ \cite{PUGiombi:2014xxa,PUFei:2015oha}
 \PUes{PUFIsing}{
  \frac{F_\text{Ising} }{F_S} \approx 0.976 \,.
 }
The $F$-coefficient of the 3-d Ising CFT is thus fairly close to that of a free scalar field, but slightly smaller than the latter as required by the $F$-theorem.  That the ratio \eqref{PUFIsing} is close to $1$ indicates that the 3-d Ising CFT is in some sense fairly close to the free scalar CFT in the space of all CFTs\@.  For comparison, the scaling dimension of $\phi$ in the Ising CFT is $\Delta_\phi^\text{Ising} \approx 0.518$, which is fairly close to the free scalar value $\Delta_\phi^\text{free} = 1/2$, but the scaling dimension of $\phi^2$ is rather different being $\Delta_{\phi^2}^\text{Ising} \approx 1.412$ in the Ising CFT and $\Delta_{\phi^2}^\text{free} \approx 1$ in the free theory.  As another comparison, the two-point function of the canonically normalized stress energy tensor $T_{\mu\nu}$ obeys $\langle T_{\mu\nu}(x) T_{\rho\sigma}(0) \rangle_\text{Ising} \approx 0.947 \, \langle T_{\mu\nu}(x) T_{\rho\sigma}(0) \rangle_\text{free} $, so again this quantity in the interacting theory is very close to the free field result.

It is worth noting that Refs.~\cite{PUGiombi:2014xxa,PUFei:2015oha,PUGiombi:2015haa} also conjectured that $\tilde F^{(d)}$ may obey a generalized version of the $F$-theorem as well as of the $F$-maximization principle to be discussed in Section~\ref{PUFMAXIMIZATION}.  (See also~\cite{PUMinahan:2015any}.)

\subsection{$F$ in theories with holographic duals}
\label{PUSUGRA}

Another class of CFTs where one can calculate the $F$-coefficient approximately are CFTs with holographic duals.  Quite generally, suppose we have such a CFT whose vacuum is dual to the $AdS_4$ solution of the classical equations of motion of an effective 4-d two-derivative gravity (or supergravity) theory with negative cosmological constant.  Let $L$ be the radius of curvature of $AdS_4$.  In Einstein frame, the two-derivative classical Euclidean action can be written as
 \PUes{PUGravityAction}{
  S = \frac{1}{16 \pi G_4} \int d^4 x \, \sqrt{g} \left(-R - \frac{6}{L^2} + {\cal L}_\text{matter} \right) \,,
 }
where $G_4$ is the 4-d Newton constant, and ${\cal L}_\text{matter}$ is the matter Lagrangian that we assume vanishes on the (Euclidean) $AdS_4$ solution of the equations of motion.  

In the regime where classical two-derivative supergravity is a good approximation, the $F$-coefficient of the dual field theory is simply equal to the regularized value of the Euclidean on-shell action.  The Euclidean continuation of $AdS_4$ is the hyperbolic space $\BH^4$, whose line element is
 \PUes{PUMetricH4}{
  ds^2 = L^2 (d r^2 + \sinh^2 r \, ds_{S^3}^2 ) \,,
 }
where $ds_{S^3}^2$ is the metric on a round three-sphere of unit curvature radius.  The Ricci scalar of \eqref{PUMetricH4} is $R =- 12/L^2$, and so plugging \eqref{PUMetricH4} into \eqref{PUGravityAction}, we can write the on-shell action as
 \PUes{PUOnShell}{
  F = \frac{3}{8 \pi G_4 L^2} \Vol(\BH^4)   \,.
 }
The volume of $\BH^4$ is of course infinite, but it can be regularized by imposing a cutoff that obeys the $SO(4)$ symmetry:
 \PUes{PUVolH4}{
  \Vol(\BH^4) = L^4 \Vol(S^3) \int_0^{\log \Lambda} dr\, \sinh^3 r = L^4 \Vol(S^3) \left[\frac{\Lambda^3}{24} - \frac{3 \Lambda}{8} + \frac 23 + \ldots \right] \,.
 }
After throwing away the power divergences in $\Lambda$ and using $\Vol(S^3) = 2 \pi^2$, we obtain $\Vol(\BH^4) = \frac 43 \pi^2 L^4$.  From \eqref{PUOnShell}, we deduce that \cite{PUHerzog:2010hf,PUJafferis:2011zi}
 \PUes{PUOnShellFinal}{
  F = \frac{\pi L^2}{2 G_4} 
 }
in the classical gravity approximation.  To find an explicit expression for $F$ in terms of field theory data, one has to use the AdS/CFT dictionary on a case-by-case basis to relate $L^2 / G_4$ to quantities defined in the field theory.

An interesting class of examples of CFTs with $AdS_4$ duals are the infrared limits of the effective theories on $N$ coincident M2-branes placed at the tip of a Calabi-Yau cone.  If $Y_7$ is the seven-dimensional Sasaki-Einstein manifold that is the base of this cone, then these CFTs are dual to the $AdS_4 \times Y_7$ solution of eleven-dimensional supergravity with $N$ units of seven-form flux threading $Y_7$.  It can be shown that in this case \cite{PUHerzog:2010hf}
 \PUes{PUFVolY}{
  F = N^{3/2} \sqrt{ \frac{2 \pi^6}{27 \Vol(Y_7)}}  \,,
 }
to leading order at large $N$.

That in the theories on $N$ coincident M2-branes the number of degrees of freedom scales as $N^{3/2}$ at large $N$ was first noticed by Klebanov and Tseytlin \cite{PUKlebanov:1996un} in the computation of the thermal free energy.  A field theory explanation of the $N^{3/2}$ scaling of the $F$-coefficient was provided with the help of supersymmetric localization, as will be reviewed in the next section.

\subsection{$F$ in supersymmetric theories}
\label{PUSUPERSYMMETRIC}

Besides free theories, the $F$-coefficient can also be computed exactly in superconformal field theories (SCFTs) that have Lagrangian descriptions, where one can use the technique of supersymmetric localization.  The idea \cite{PUWitten:1988ze} of supersymmetric localization is as follows.  In a supersymmetric theory, the value of a functional integral does not change if we add a term to the action that is $Q$-exact, $Q$ being a supercharge under which the action is invariant.  Moreover, if this $Q$-exact term has a positive-definite bosonic part, then adding it with a large coefficient allows for the evaluation of the functional integral in the saddle point approximation, which in this limit becomes exact.  The saddles on which the functional integral localizes are those on which the $Q$-exact term vanishes.  In favorable circumstances, as is the case of ${\cal N} \geq 2$ theories on $S^3$, the localization locus is finite dimensional, so the partition function can be expressed as a finite dimensional integral.

Building on the work of \cite{PUPestun:2007rz}, the technique of supersymmetric localization was first applied to three-dimensional SCFTs on $S^3$ in \cite{PUKapustin:2009kz}.  Ref.~\cite{PUKapustin:2009kz} focused on the case of SCFTs with Lagrangian descriptions of the type constructed in \cite{PUSchwarz:2004yj, PUGaiotto:2007qi} whose Lagrangians are invariant under superconformal transformations.   All theories of this type that preserve ${\cal N} \geq 3$ supersymmetry have this property.  These results were then further generalized to ${\cal N} = 2$ SCFTs that are embedded as deep IR limits of RG flows triggered by relevant superpotential interactions \cite{PUJafferis:2010un}.  As will be explained in the next section, in these ${\cal N} = 2$ examples one also has to supplement the supersymmetric localization with the technique of $F$-maximization \cite{PUJafferis:2010un, PUJafferis:2011zi, PUClosset:2012vg}.  It is not currently known how to calculate the $F$-coefficient of SCFTs with only ${\cal N} = 1$ supersymmetry using similar methods.   In the remainder of this section, let us focus on ${\cal N} \geq 3$ SCFTs with exactly marginal Lagrangians as in \cite{PUSchwarz:2004yj, PUGaiotto:2007qi}, and defer the discussion of ${\cal N} = 2$ SCFTs to Section~\ref{PUFMAXIMIZATION}.  

The field content of these ${\cal N} \geq 3$ SCFTs can be described in terms of vector multiplets and hypermultiplets.  While we provide explicit Lagrangians in an ${\cal N} =2$ notation  and more details of the supersymmetric localization computation in the next section, let us now simply state the results in the case of ${\cal N} \geq 3$ SCFTs and provide a few examples.   Let our ${\cal N} = 3$ SCFT have gauge group $G$, written as a product of simple factors,   $G = G^{1} \times G^{2} \times \cdots \times G^{n}$, and with the Chern-Simons level of each factor denoted by $k_a$.  The matter content consists of hypermultiplets ${\cal H}_i$ transforming in representations $R_i$ of the gauge group.   For such a theory, it is explained in detail in~\volcite{WI} that upon using the technique of supersymmetric localization the partition function takes the form
 \PUes{PUZLoc}{
  Z = \frac{1}{\abs{\cal W}} \int \prod_\text{Cartan} d \sigma \exp\left[\frac{i}{4 \pi} \tr_k \sigma^2  \right] \frac{\det_\text{Ad} \left( 2 \sinh (\pi \sigma) \right)}{ \prod_{\substack{\text{hypers} \\ \text{in rep $R_i$}}} \det_{R_i} \left( 2 \cosh (\pi \sigma)  \right) } \,.
 }
Here, $\sigma$ is an element of the Lie algebra that lies along the Cartan subalgebra;  in particular, it can be written as $\sigma = \sigma_a h_a$.  The determinant of $f(\sigma)$ in a representation $R$ with weights $w_a^i$, with $i$ ranging from $1$ to the $\dim R$, is defined as
 \PUes{PUDetDef}{
  \text{det}_R\, f(\sigma) \equiv \prod_i f(\sum_a \sigma_a w_a^i) \,.
 }
The determinant in the adjoint in the numerator of \eqref{PUZLoc} is defined as \eqref{PUDetDef} but without including the zero weights (i.e.~the Cartan elements) in the product.  Lastly, $\abs{{\cal W}}$ is the order of the Weyl group of the gauge group. 

\subsubsection{Examples}
\label{PUSUSYEXAMPLES}

As a first example, let us consider ${\cal N} = 4$ $U(1)$ super-QED with $N$ hypermultiplets of unit gauge charge.  ${\cal N} = 4$ supersymmetry requires $k=0$ in this case.  From \eqref{PUZLoc}, we have \cite{PUKlebanov:2011td}
 \PUes{PUZU1}{
  Z = \frac{1}{2^N} \int_{-\infty}^\infty \frac{d \lambda}{\cosh^N (\pi \lambda)} = \frac{\Gamma\left( \frac N2 \right)}{2^N \sqrt{\pi} \Gamma \left( \frac{N+1}{2} \right) } \,.
 }
Extracting $F = - \log \abs{Z}$ and expanding it at large $N$ we obtain
 \PUes{PUFLargeNU1}{
  F = N \log 2 + \frac 12 \log \left( \frac{N \pi}{2} \right) - \frac{1}{4 N} + \frac{1}{24 N^3} + \cdots \,.
 }
Quite nicely, this expression matches the first few terms of \eqref{PUFComplicated}, as can be easily checked after noting that $N$ hypermultiplets contain $N_b = 4N$ real scalar fields and $N_f = 2N$ Dirac fermions.  In addition, when $N=1$, \eqref{PUZU1} gives $Z = \frac 12$, or $F = \log 2$, which is the same value of $F$ as for a free hypermultiplet---see \eqref{PUGotFN2}.  Indeed, ${\cal N} = 4$ SQED with one charged hypermultiplet is known to be mirror dual to a free (twisted) hypermultiplet \cite{PUIntriligator:1996ex}.

Another application of \eqref{PUZLoc} is to SCFTs with holographic duals  \cite{PUDrukker:2010nc, PUHerzog:2010hf, PUSantamaria:2010dm, PUGulotta:2011si, PUMarino:2011eh}, where it provides a field theory explanation for the effective number of degrees of freedom of theories on coincident M2-branes mentioned in Section~\ref{PUSUGRA}.  One of the simplest such examples is that of $N$ M2-brane placed at a $\mathbb{C}^4 / \BZ_k$ singularity, where the space $Y_7$ appearing in \eqref{PUFVolY} is a freely acting $\BZ_k$ orbifold of the 7-sphere, $Y_7 = S^7 / \BZ_k$.  The dual field theory is an ${\cal N} = 6$ theory, which in ${\cal N} = 3$ language is a $U(N) \times U(N)$ gauge theory with Chern-Simons levels $k$ and $-k$ for the two gauge groups, and with matter given by two bi-fundamental hypermultiplets \cite{PUAharony:2008ug}.\footnote{In ${\cal N} = 4$ notation, the matter fields consist of a bi-fundamental hypermultiplet and a bi-fundamental twisted hypermultiplet.  The distinction between hypermultiplets and twisted hypermultiplets is lost when restricting to ${\cal N} =3$ supersymmetry.}  Eq.~\eqref{PUZLoc} takes the form
 \PUes{PUZABJM}{
  Z = \frac{1}{(N!)^2} \int \left(\prod_{i=1}^N  d\lambda_i d\tilde \lambda_i \right)
   \frac{\prod_{i<j}  \left( 4 \sinh \left[ \pi (\lambda_i - \lambda_j) \right] 
     \sinh \left[ \pi (\tilde \lambda_i - \tilde \lambda_j) \right]  \right)^2}{\prod_{i, j} \left(
        2 \cosh \left[ \pi (\lambda_i -\tilde \lambda_j )\right] \right)^2}  
         e^{i \pi k \sum_i (\lambda_i^2 - \tilde \lambda_i^2)} \,.
 }
This integral can be solved through a variety of methods \cite{PUDrukker:2010nc, PUHerzog:2010hf, PUMarino:2011eh}.  At large $N$ and fixed $k$, it gives
 \PUes{PUFreeABJM}{
  F = \frac{\pi \sqrt{2}}{3} k^{1/2} N^{3/2} + O(N^{1/2}) \,,
 }
thus reproducing the supergravity expectation \eqref{PUFVolY} after using $\Vol(S^7 / \BZ_k) = \pi^4 / (3k)$.   (See also~\volcite{MA} for a calculation of $F$ in a `t Hooft-like limit where $N$ is taken to infinity while keeping $N/k$ fixed.)  This field theory calculation of $F$ provides, in this example, a derivation of the $N^{3/2}$ scaling behavior of the number of degrees of freedom on $N$ coincident M2-branes without relying on the dual supergravity description. Various other generalizations to other 3-d theories with supergravity duals were considered in \cite{PUMartelli:2011qj, PUJafferis:2011zi, PUCheon:2011vi, PUGulotta:2011si,PUHerzog:2010hf,  PUSantamaria:2010dm, PUGabella:2011sg, PUGulotta:2011vp, PUGulotta:2012yd, PUCrichigno:2012sk,PUAmariti:2012tj,PUAssel:2012cp,PUGuarino:2015jca}.   They are rather stringent tests of the corresponding holographic dualities.

\section{${\cal N} = 2$ SCFTs and $F$-maximization}
\label{PUFMAXIMIZATION}

In generalizing the technique of supersymmetric localization to ${\cal N} = 2$ SCFTs with Lagrangian descriptions one faces the following challenge.  In flat space, most of these SCFTs are described as IR fixed points of non-trivial RG flows, and one simply cannot write down a Lagrangian for these IR fixed points that is superconformal, as was the case for the ${\cal N} \geq 3$ SCFTs.  Consequently, one cannot use the stereographic projection map to put a generic ${\cal N} = 2$ SCFT on $S^3$ directly, without having to rely on its definition as the IR limit of an RG trajectory.  In general, mapping the RG trajectory from $\BR^3$ to $S^3$ is ambiguous because there are many curvature couplings one can add in this process.

The ambiguities in mapping from $\BR^3$ to $S^3$ can be fixed and one can generalize the technique of supersymmetric localization if the RG trajectory on $\BR^3$ preserves a $U(1)$ R-symmetry.  If flavor symmetries are also present throughout the RG trajectory, then this $U(1)_R$ symmetry may not be unique, because any linear combination of a given $U(1)$ R-symmetry and any of the $U(1)$ flavor symmetries is also an R-symmetry.

For any choice of $U(1)_R$ symmetry preserved by the RG flow, it is possible to map the RG flow from $\BR^3$ to $S^3$ uniquely by requiring that the curvature couplings are such that the RG flow on $S^3$ preserves an $SU(2|1) \times SU(2)$ symmetry, whose bosonic part consists of the $SO(4) = SU(2) \times SU(2)$ isometry group of $S^3$ as well as the $U(1)_R$ symmetry mentioned above. (There is only a discrete choice as to which $SU(2)$ subgroup of $SO(4)$ is part of $SU(2|1)$.) In the case of Chern-Simons matter theories, explicit Lagrangians invariant under $SU(2|1) \times SU(2)$ were constructed in \cite{PUJafferis:2010un, PUHama:2010av}, and a more systematic approach based on coupling the flat space theory to a background supergravity multiplet was developed in \cite{PUFestuccia:2011ws, PUClosset:2012vp, PUClosset:2012vg}.  (See also \volcite{DU}.)

\subsection{Non-conformal theories on $S^3$}
\label{PUACTION}

To be concrete, let us briefly review the construction of these $SU(2|1) \times SU(2)$ invariant actions on $S^3$.\footnote{\label{Conventions}We use the notation in \cite{PUFreedman:2013ryh}.  In particular, we take the Euclidean gamma matrices to be given by the Pauli matrices, $\gamma^i = \sigma^i$, where $i$ is a frame index that is raised and lowered with the flat 3-d Euclidean metric. We use the frame $e^i_\mu$ given by the left-invariant one-forms.  In this frame the spacetime covariant derivative on a spinor $\psi$ can be written as $\nabla_\mu \psi = (\partial_\mu + \frac{i}{2} \gamma_\mu)\psi$, where $\gamma_\mu = \gamma_i e^i_\mu$.} (See also \volcite{WI}.) As mentioned above, there is a discrete choice in constructing such actions corresponding to which $SU(2)$ subgroup of the $SO(4) \cong SU(2)_L \times SU(2)_R$ isometry group of $S^3$ is the one contained in $SU(2|1)$.  This choice is manifested in which Killing spinors are chosen as supersymmetry transformation parameters.  We choose to construct actions invariant under supersymmetry transformations with $\epsilon$, $\tilde \epsilon$ obeying
 \PUes{PUKillingSpinors}{
  \nabla_\mu \epsilon = \frac{i}{2 a} \sigma_\mu \epsilon \,, \qquad
    \nabla_\mu \tilde \epsilon = \frac{i}{2 a} \sigma_\mu \tilde \epsilon \,,
 }
where $a$ is the radius of the sphere.  These spinors are invariant under $SU(2)_L$, but transform under $SU(2)_R$ as a doublet, provided that we work in a frame given by $SU(2)_L$-invariant one forms on $S^3$---see footnote~\ref{Conventions}.  Therefore, if one chooses the transformation parameters to obey \eqref{PUKillingSpinors}, the theories we will consider on $S^3$ are invariant under $SU(2|1)_R \times SU(2)_L$, where $SU(2)_R$ is contained in $SU(2|1)_R$.  Similar actions can be obtained using $SU(2)_R$-invariant spinors by formally sending $a \to -a$ in \eqref{PUKillingSpinors} and in all the formulas below.

Just as in the case of ${\cal N} = 1$ supersymmetric theories in four dimensions, the explicit actions can be constructed from vector multiplets and chiral multiplets as follows.  Let us write the gauge group as a product of simple factors,
 \PUes{PUGTotal}{
  G^{1} \times G^{2} \times \cdots \times G^{n} \,,
 }
and denote the vector multiplet associated to each simple factor by ${\cal V}^{a} = (A_\mu^{a}, \lambda^{a}, \sigma^{a}, D^{a})$, $a = 1, \ldots, n$, where $A_\mu^{a}$ is a gauge field, $\lambda^{a}$ is the gaugino, and $\sigma^{a}$ and $D^{a}$ are scalar fields, all transforming in the adjoint representation of $G^{a}$.  A chiral multiplet $\Phi^{i} = (Z^{i}, \chi^{i}, F^{i})$, with $Z^{i}$ and $F^{i}$ scalar fields and $\chi^{i}$ a two-component spinor, can in general transform in any representation of the product gauge group \eqref{PUGTotal}.  In Lorentzian signature $A_\mu^{a}$, $\sigma^{a}$, and $D^{a}$ would be real, while the other fields would be complex.  Let us denote $\tilde \lambda^{a} = i \sigma_2 \lambda^{a*}$,  $\tilde \chi^{i} = i \sigma_2 \chi^{i*}$, $\tilde Z^{i} = Z^{i*}$, and $\tilde F^{i} = F^{i*}$.   In Euclidean signature, we should allow  $A_\mu^{a}$, $\sigma^{a}$, and $D^{a}$ to be complex and thus formally treat the tilde'd fields as independent from the untilded'd ones.

We would like to construct an action on $S^3$ that is invariant under $SU(2|1) \times SU(2)$.  The transformations of the fields under the fermionic part of this superalgebra, with independent supersymmetry parameters $\epsilon$ and $\tilde \epsilon$ obeying \eqref{PUKillingSpinors}, can be realized on the fields as
\PUes{PUVectorTransf}{
  \delta A_\mu &=  -\frac i2  \epsilon^T (i \sigma_2) \sigma_\mu \tilde \lambda  -\frac i2  \tilde \epsilon^T (i \sigma_2) \sigma_\mu \lambda \,, \\
  \delta \sigma &= - \frac 12\epsilon^T (i \sigma_2) \tilde \lambda + \frac 12 \tilde \epsilon^T (i \sigma_2) \lambda \,, \\
  \delta D &=  -\frac 12 \epsilon^T (i \sigma_2) \left( \sigma^\mu \nabla_\mu \tilde \lambda - \frac{i}{2a} \tilde \lambda \right) 
   + \frac 12 \tilde \epsilon^T (i \sigma_2) \left(\sigma^\mu \nabla_\mu \lambda - \frac{i}{2a} \lambda \right) \,, \\
  \delta \lambda &= \left(\frac 12 \sigma^{\mu\nu} F_{\mu\nu} + i \sigma^\mu \partial_\mu \sigma + i D - \frac 1a \sigma \right) \epsilon  \,, \\
  \delta \tilde \lambda &= \left(\frac 12 \sigma^{\mu\nu} F_{\mu\nu} - i \sigma^\mu \partial_\mu \sigma - i D + \frac 1a \sigma \right) \tilde \epsilon 
 }
for a vector multiplet, and
 \PUes{PUChiralTransf}{
  \delta Z &= i \epsilon^T (i \sigma_2) \chi \,, \\
  \delta \tilde Z &= i \tilde \epsilon^T (i \sigma_2) \tilde \chi \,, \\
  \delta F &= \tilde \epsilon^T (i \sigma_2) \left(i \sigma^\mu \nabla_\mu \chi + i  \left( \sigma -  i \frac{r - 1/2}{a} \right) \chi - \tilde \lambda Z \right) \,,\\
  \delta \tilde F &= \epsilon^T (i \sigma_2) \left(i \sigma^\mu \nabla_\mu \tilde \chi + i  \left( \sigma -  i \frac{r - 1/2}{a} \right) \tilde \chi + \lambda \tilde Z \right) \,, \\
  \delta \chi &= F \epsilon + \left(\sigma^\mu \nabla_\mu Z -  \sigma Z + \frac{i r}{a} Z \right) \tilde \epsilon \,, \\
  \delta \tilde \chi &= \left(\sigma^\mu \nabla_\mu \tilde Z -  \sigma \tilde Z + \frac{i r}{a} \tilde Z \right) \epsilon + \tilde F \tilde \epsilon 
 }
for a chiral multiplet.    Indeed, in the $a \to \infty$ limit, the supersymmetry transformations rules listed above are the usual ones from flat space.  The $1/a$ corrections are precisely what is needed in order to realize the commutation rules of the $SU(2|1)$ algebra: writing the supersymmetry variations in terms of the supercharges,
 \PUes{PUdeltaToQ}{
  \delta = i \epsilon^T (i \sigma_2) Q +  i \tilde \epsilon^T (i \sigma_2) \tilde Q \,,
 } 
it is straightforward to check that \eqref{PUVectorTransf}--\eqref{PUChiralTransf} imply 
 \PUes{PUSUSYAlgebra}{
  \{ Q, \tilde Q^T i \sigma_2 \} = \sigma^\mu J_i + i \sigma + \frac{1}{a} R \,,
 }
where $J_i$ is an $SU(2)_R$ isometry, and $R$ is the R-charge.  For a chiral multiplet $(Z, \chi, F)$, the R-charges are $(r,  r-1,  r-2)$.  The anti-chiral multiplet $(\tilde Z, \tilde \chi, \tilde F)$ has opposite R-charges.

The total action on $S^3$ preserving $SU(2|1) \times SU(2)$, with a given choice of $U(1)_R$ R-symmetry contained in $SU(2|1)$, can be written as a sum of four terms: 
 \PUes{PUSTotal}{
  S = S_\text{CS} + S_\text{kin} + S_\text{superpot}  + S_\text{top} \,.
 }
The first term corresponds to a possible Chern-Simons interaction term with level $k^{a}$ for the gauge group factor $G^{a}$.  To avoid clutter, let us write down the Chern-Simons term for a vector multiplet ${\cal V} = (A_\mu, \lambda, \sigma, D)$ with level $k$, thus dropping the superscript $a$:
 \PUes{PUCS}{
  S_\text{CS}[{\cal V}; k] = \frac{i k}{4 \pi} \int d^3 x \tr \left[ \epsilon^{\mu \nu \rho} \left( A_\mu \partial_\nu A_\rho - \frac {2i}3 A_\mu [A_\nu, A_\rho] \right) - \sqrt{g} (\tilde \lambda^{T} (i \sigma_2) \lambda + 2 i \sigma D) \right] \,.
 }
This is simply the supersymmetrized version of the Chern-Simons term in \eqref{PUActionCS}.  The second term in \eqref{PUSTotal} includes the kinetic terms for the chiral multiplets.  The kinetic term for a chiral multiplet $\Phi = (Z, \chi, F)$ with $U(1)_R$ charge $r$ is
 \PUes{PUSkin}{
  S_\text{kin}[\Phi, r] &= \int d^3 x \sqrt{g} \tr \Biggl(\nabla^\mu \tilde Z \nabla_\mu Z
   +  \tilde Z  \left( \sigma -  i \frac{r - 1/2}{a} \right)^2 Z + i \tilde \chi^T (i \sigma_2) \sigma^\mu \nabla_\mu \chi 
   \\ &{}+ i \tilde \chi^T (i \sigma_2) \left( \sigma -  i \frac{r - 1/2}{a} \right) \chi - \tilde F F \\
   &{}+ \lambda^T (i \sigma_2) \tilde Z \chi + \tilde \chi^T (i \sigma_2) Z \tilde \lambda - \left( D - \frac{r - 1/2}{a^2} \right) \tilde Z Z + \frac{3}{4 a^2} \tilde Z Z \Biggr) \,.
 }
Here, $\nabla_\mu$ includes the gauge covariant derivative, namely $\nabla_\mu \chi = (\partial_\mu  + \frac{i}2 \gamma_\mu - i A_\mu) \chi$, $\nabla_\mu Z = (\partial_\mu - i A_\mu) Z$, etc. (See footnote~\ref{Conventions} for our frame and gamma matrix conventions.)  The total Chern-Simons and kinetic terms are
 \PUes{PUSkinTotal}{
  S_\text{CS} = \sum_{a=1}^n S_\text{CS}[{\cal V}^{a};{k_{a}}] \,, 
   \qquad S_\text{kin} = \sum_i S_\text{kin}[\Phi^i, r_i] \,,
 }
where the sum is over all chiral multiplets $\Phi^i$ with R-charge $r_i$.  These terms are invariant under \eqref{PUVectorTransf}--\eqref{PUChiralTransf} for any choices of $r_i$.  

The third term in \eqref{PUSTotal}, $S_\text{superpot}$, corresponds to superpotential interactions and is given by
 \PUes{PUSW}{
  S_\text{superpot} = \int d^3 x \sqrt{g} \left[ F^{i} W_i  + \frac{1}{2} W_{ij}
   \chi^{i} \sigma_2 \chi^{j} + \tilde F^{i} \tilde W_i  + \frac{1}{2} \tilde W_{ij}
   \tilde \chi^{i} \sigma_2 \tilde \chi^{j} \right] \,,
 }
where $W = W(Z_i)$, $\tilde W = \bar W(\tilde Z_i)$, $W_i = \frac{\partial W(Z_i)}{\partial Z^i}$, etc.  This term is invariant under \eqref{PUVectorTransf}--\eqref{PUChiralTransf} only if the superpotential satisfies
 \PUes{PUWRelation}{
  (r_i - 2) W_i + \sum_j r_j W_{ij} Z^j = 0 \,.
 }
This condition is obeyed if $W$ is a sum of monomials in the $Z^i$ with the property that the sum of the R-charges of each monomial equals two.  It thus follows that the trial R-charges $r_i$ are not arbitrary, but that they are constrained by the condition that the superpotential should have R-charge two.

It is worth noting that if one sends $r_i \to r_i + i m_i a$ in the action above, the parameters $m_i$ are nothing but ``real masses'' for the chiral multiplets.  The $S^3$ action is therefore holomorphic in $r_i + i m_i$, a property first noticed in \cite{PUJafferis:2010un} and later explained in \cite{PUClosset:2012vg}.  The explanation of this holomorphy provided in \cite{PUClosset:2012vg} is that a change in $r_i + i m_i$ can be realized by coupling one of the conserved currents of the theory to a background vector multiplet, and then giving supersymmetry-preserving expectation values to the scalars in the background vector multiplet.  See also \volcite{DU}.  We will return to this point of view shortly.

The last term in \eqref{PUSTotal} is more subtle, but can become important in a gauge theory where the gauge group has a $U(1)$ or $U(N)$ factor with no Chern-Simons interactions.  (In other situations it may be ignored.)  For every such factor, the contribution to the last term in \eqref{PUSTotal} is an FI term
 \PUes{PUStop}{
  S_\text{top}[{\cal V}; r_\text{top}] = \frac{r_\text{top}}{2 \pi a} \int d^3x \sqrt{g}\, \tr \left(i D + \frac{\sigma}{a} \right)\,,
 }
with parameter $r_\text{top}$; the total term is
 \PUes{PUStopTotal}{
  S_\text{top} = \sum_a S_\text{top}[{\cal V}; r^a_\text{top}] \,.
 }
As written in \eqref{PUStop}, real coefficients $r_\text{top}^a$ correspond to pure imaginary FI parameters.  As we will now explain, the coefficients $r^a_\text{top}$ are in some sense on the same footing as the R-charges $r_i$ in \eqref{PUSkinTotal} in that they are part of the definition of which $U(1)_R$ symmetry was used to place the theory on $S^3$.

Indeed, for a $U(1)$ or $U(N)$ gauge group factor, one can construct the conserved current
 \PUes{PUjtop}{
  j^\mu_\text{top} = \frac{1}{4 \pi} \epsilon^{\mu\nu \rho} \tr F_{\nu\rho} \,,
 }
usually referred to as a topological current because its conservation follows simply from the Bianchi identity obeyed by $F_{\mu\nu}$.   No operators constructed as polynomials in the matter fields alone are charged under $j^\mu_\text{top}$.  The only operators charged under $j^\mu_\text{top}$ can be monopole operators \cite{PUBorokhov:2002cg,PUBorokhov:2002ib}, which are defined through certain boundary conditions that the gauge field as well as the other fields in the ${\cal N} = 2$ vector multiplet should satisfy close to the insertion point.  These operators can only carry integer units of topological charge $q_\text{top} = \int d^2 x j^0_\text{top} \in \BZ$.  In a supersymmetric theory, these operators may carry R-charge as well as other flavor charges.  For instance, a half-BPS monopole operators ${\cal M}$ of charge $q_\text{top}$ has R-charge
 \PUes{PURchargeMon}{
  \gamma (\abs{q_\text{top}}) + r_\text{top} q_\text{top}
 }
where $\gamma (\abs{q_\text{top}}) $ can be computed at one-loop and depends only on the absolute value of $q_\text{top}$  \cite{PUBenini:2009qs, PUJafferis:2009th}, and $r_\text{top}$ is the parameter in \eqref{PUStop}.  Thus, in specifying the $U(1)_R$ symmetry used to place a Chern-Simons matter theory on $S^3$ requires specifying both the R-charges $r_i$ of the chiral matter fields and the R-charge parameters $r_\text{top}$ of the chiral monopole operators.

Just as if one sends $r_i \to r_i + i m_i a$ can introduce a real mass deformation of the $S^3$ theory, sending $r_\text{top} \to r_\text{top} + i \zeta$ introduces an FI parameter $\zeta$.  The $S^3$ partition function is thus holomorphic in $r_\text{top} + i \zeta$, as can be understood from the fact that $r_\text{top} + i \zeta$ arises as the expectation value of a background vector multiplet that couples to the conserved current multiplet that contains the topological current \eqref{PUjtop}.  See \cite{PUClosset:2012vg} and \volcite{DU}.

The explicit construction of $SU(2|1) \times SU(2)$-invariant theories on $S^3$ presented above can be rephrased in a more abstract language \cite{PUFestuccia:2011ws, PUClosset:2012vp, PUClosset:2012vg}.  Suppose we start with the non-conformal theory on $\BR^3$ that is believed to flow to our SCFT of interest in the IR---in other words, suppose we start with the flat space $a\to \infty$ limit of \eqref{PUSTotal} and try to deduce the various terms proportional to $1/a$ and $1/a^2$ in \eqref{PUVectorTransf}--\eqref{PUStop}.  We choose a $U(1)_R$ symmetry preserved by this non-conformal theory in flat space.  One can find a unique super-multiplet ${\cal R}_\mu$ that contains the $U(1)_R$ current, the stress-energy tensor, the supersymmetry current, a conserved current corresponding to the central charge of the supersymmetry algebra, and a string current.  This multiplet can then be coupled to a background supergravity multiplet.  The supergravity multiplet contains the metric $g_{\mu\nu}$, the gravitino, two Abelian gauge fields, and a two-form gauge field.  The various terms proportional to $1/a$ and $1/a^2$ in \eqref{PUVectorTransf}--\eqref{PUStop} correspond to non-vanishing background values for the fields in the ${\cal H}_\mu$ multiplet required in order to preserve supersymmetry.  See \cite{PUFestuccia:2011ws, PUClosset:2012vp, PUClosset:2012vg} as well as \volcite{DU} for more details.

\subsection{Supersymmetric localization}
\label{PULOCALIZE}

The $S^3$ partition function of Chern-Simons matter theories with the action \eqref{PUSTotal} can be computed using supersymmetric localization \cite{PUJafferis:2010un, PUHama:2010av} building on the work of \cite{PUKapustin:2009kz}.  The idea is that
 \PUes{PUZLocGen}{
  Z = \int e^{-S} = \int e^{ - S_t} \,, \qquad S_t \equiv S + t\, \int d^3 x \sqrt{g} \{Q, P \} \,,
 }
for some suitable operator $P$ and supercharge $Q$ such that the bosonic part of $\{Q, P \}$ is positive definite.  Taking $t \to \infty$ in \eqref{PUZLocGen}, one can evaluate this expression in the saddle point approximation by considering quadratic fluctuations around the configurations where $ \{Q, P \} = 0$:
 \PUes{PUZSaddle}{
  Z = e^{-S\big|_{\{Q, P\} = 0}} \times \text{(1-loop det)} \,.
 }

The choice
 \PUes{PUPChoice}{
  P = \{Q, \lambda\}^\dagger \lambda + \{Q, \chi\}^\dagger \chi  +\tilde \chi \{Q, \tilde \chi\}^\dagger
 }
obeys the properties mentioned above.  Moreover, $\{Q, P \} = 0$ implies that all the fields vanish except for $\sigma^a$ and $D^a$, which are required to take constant values related by
 \PUes{PUDsigmaRel}{
  \{Q, P\} = 0 \qquad \Longrightarrow \qquad D^a = -\frac{i \sigma^a}{a} \,.
 }
On this configuration, only the first and last terms in \eqref{PUSTotal} have a non-zero contribution.  For a vector multiplet ${\cal V}$ with Chern-Simons level $k$ and imaginary FI parameter $r_\text{loc}$, this classical contribution is 
 \PUes{PUSLoc}{
  S_\text{classical}[{\cal V}; k; r_\text{top}] =  -\int d^3x \sqrt{g} \left[ \frac{i k}{2 \pi a}   \tr   \sigma^2 - \frac{r_\text{top}}{\pi a^2}  \tr  \sigma \right] = -i \pi a^2  k   \tr \sigma^2 - 2 \pi r_\text{top} a \tr \sigma  \,,
 }
where we used the fact that the volume of a three-sphere of radius $a$ is $2 \pi^2 a^3$.   We will set the radius of the sphere to $a=1$ from now on.

The computation of the one-loop determinants is tedious and performed in detail in \volcite{WI}.  Here, we will only list the results.    The one-loop determinant for a vector multiplet ${\cal V}$ is
 \PUes{PUDetV}{
  Z_\text{1-loop}[{\cal V}] = \text{det}_\text{Ad} (2 \sinh (\pi \sigma) ) \,,
 } 
where the determinant in the adjoint representation is defined as in \eqref{PUDetDef}, but without including the zero weights in the product.  For a chiral multiplet $\Phi$ of R-charge $r$ transforming in representation $R$ of the gauge group, the one-loop determinant is
 \PUes{PUDetC}{
  Z_\text{1-loop}[\Phi; r]
   = \text{det}_R e^{\ell(1 - r + i \sigma)} \,,
 }
where the determinant in representation $R$ was defined in \eqref{PUDetDef}, and the function $\ell(z)$ is defined by $\ell'(z) = -\pi z \cot \pi z$ and $\ell(0) = 0$.  Explicitly,
 \PUes{PUellExplicit}{
  \ell(z) = - z \log \left( 1 - e^{2 \pi i z} \right) + \frac i2 \left( \pi z^2 + \frac 1 \pi \text{Li}_2 \left( e^{2 \pi i z} \right) \right)  
   - \frac{i \pi}{12} \,.
 }

Combining these expressions, we can write the partition function as \cite{PUJafferis:2010un}
 \PUes{PUGotZLoc}{
  Z = \frac{1}{\abs{\cal W}} \int_\text{Cartan} d \sigma  
   \prod_a  \left[e^{ i \pi  k_a   \tr (\sigma^a)^2 - 2 \pi r_\text{top}^a \tr \sigma^a }  \text{det}_\text{Ad} (2 \sinh (\pi \sigma^a) ) \right]
    \prod_i \text{det}_{R_i} e^{\ell(1 - r_i + i \sigma)} \,.
 }
Here $\abs{\cal W}$ is the order of the Weyl group of the gauge group.

\subsection{$F$-maximization}
\label{PUMAXIMIZE}

In Section~\ref{PUACTION} we started with a non-conformal theory on $\BR^3$ preserving a $U(1)_R$ R-symmetry $R$ and used this R-symmetry to couple this theory to curvature in a supersymmetric way.  We wrote down an action on $S^3$ preserving $SU(2|1) \times SU(2)$ where $U(1)_R$ appears in the $SU(2|1)$ algebra.  In Section~\ref{PULOCALIZE} we explained how to evaluate the $S^3$ partition function $Z$ for this theory.  We now discuss in more detail the freedom one has in this construction, and how one can determine the $U(1)_R$ symmetry that is part of the superconformal algebra of the IR SCFT on $\BR^3$.

As mentioned above, if in addition to a $U(1)_R$ symmetry, the non-conformal theory on $\BR^3$ also preserves Abelian flavor symmetry, then there is no unique choice for the $U(1)_R$ symmetry that can be used in the above construction.  Indeed, if the $U(1)_R$ current for some (canonical) choice of the $U(1)_R$ symmetry is $j_\mu^{(R_0)}$, and the Abelian flavor currents are $j_\mu^I$, where $I = 1, \ldots, F$, $F$ being the number of flavor symmetries, then 
 \PUes{PUjR}{
  j_\mu^R = j_\mu^{R_0} + \sum_{I=1}^F t_I j_\mu^I \,,
 }
with arbitrary $t_I$, is also an R-symmetry current.  The flavor currents $j_\mu^I$ could be either acting on the matter fields or could be topological currents as in \eqref{PUjtop} or, more generally, they could be linear combinations of both types of terms.  Eq.~\eqref{PUjR} represents a possible mixing of the R-symmetry with the Abelian flavor symmetries.  For each choice of the $t_I$ we can construct a different $SU(2|1) \times SU(2)$-invariant theory on $S^3$, and so the localization computation in Section~\ref{PULOCALIZE} yields an $S^3$ partition function $Z(t_I)$ that depends on the $t_I$.  Indeed, for a chiral multiplet $\Phi_i $ we can consider R-charges 
 \PUes{PUrChiral}{
  r_i = r_{i0} + \sum_{I = 1}^F t_I q^I_{i} \,,
 }
where $q^I_{i}$ is the flavor charge of $\Phi_i$ under the flavor symmetry generated by $j_\mu^I$.  Upon substitution of \eqref{PUrChiral} into \eqref{PUChiralTransf}, it can be seen that $t_I q^I_{i}$ appears in the transformation rules precisely as an expectation value for the background vector multiplet ${\cal V}^\text{bg}_I$ that couples to the conserved current multiplet that contains $j_\mu^I$.  This expectation value is 
 \PUes{PUbgExpectation}{
  A_{I\mu}^\text{bg} = \lambda_{I}^\text{bg} = 0 \,, \qquad
    \sigma^\text{bg}_I = -\frac{i t_I}{a}\,, \qquad D^\text{bg}_I = - \frac{t_I}{a^2} \,.
 }
From the transformation rules of a vector multiplet \eqref{PUVectorTransf}, we see that the configuration \eqref{PUbgExpectation} is supersymmetric.  Similarly, for a monopole operator, we can consider the R-charge parameters $r_\text{top}^a$ appearing in \eqref{PUStopTotal} to be
 \PUes{PUrMonopole}{
  r_\text{top}^a = r_{\text{top}0}^a + \sum_{I=1}^F t_I q^{Ia} \,,
 }
where $q^{Ia}$ is the charge of the monopole operator under the flavor symmetry $j_\mu^I$.  (For a monopole operator to be charged under $j_\mu^I$, it must be that $j_\mu^I$ must contain a linear combination of topological symmetries.)  The term \eqref{PUStop} in the action is then precisely the coupling of the background vector multiplet in \eqref{PUbgExpectation} to the conserved current multiplet containing the topological current.

Let us recall why we placed a non-conformal theory on $S^3$ in the first place:  we wanted to learn about the SCFT that sits in the deep IR of the RG flow on $\BR^3$.  This SCFT is invariant under an unambiguous $U(1)_R$ symmetry that appears in the superconformal algebra.  Thus the superconformal $U(1)_R$ symmetry must correspond to a specific value $t_I = t_{I*}$ of the parameters $t_I$ appearing in \eqref{PUjR}.  The statement of $F$-maximization is that the function 
 \PUes{PUFDef}{
  F(t_I) \equiv - \log \abs{Z(t_I)} =  - \text{Re}\, \log Z(t_I)
 }
is locally maximized at $t_I = t_{I*}$.\footnote{\label{PUFlat}If one constructs Chern-Simons matter theories as in Section~\ref{PUACTION}, with R-charge $r_i$ for each chiral multiplet and R-charge parameters $r^a_\text{top}$ for each Abelian gauge group factor, then the $S^3$ partition function will have flat directions.  The number of flat directions is given by the number of Abelian gauge group factors.  They correspond to shifting the R-charges by any multiple of the gauge charge.  (For more details, see Section~2.3 of \cite{PUJafferis:2011zi}.)  Consequently, $F$ depends only on the R-charges of gauge invariant operators, and $F$-maximization should be performed modulo these flat directions.}   $F$-maximization is thus a procedure for determining the R-symmetry that appears in the superconformal algebra in cases where mixing with Abelian flavor symmetries is possible.  

Showing that \eqref{PUFDef} is maximized at $t_I = t_{I*}$ requires a careful analysis of contact terms and various relations required by supersymmetry.   Intuitively, from \eqref{PUbgExpectation}, we see that taking derivatives of $Z(t_I)$ with respect to $t_I$ corresponds to insertions of the integrated operator in the supermultiplet containing $j_\mu^I$ that couples to $t_I$ via \eqref{PUbgExpectation}.  Thus, derivatives of $Z(t_I)$ evaluated at $t_I = t_{I*}$ can be expressed in terms of integrated correlation functions of operators in the conserved current multiplet in the SCFT, which are all parameterized by just a few numbers.  In particular, the first derivative $\partial F / \partial t_I$ should vanish when $t_I = t_{I*}$ because it equals an integrated one-point function in a CFT\@.  The second derivative of $\partial^2 F / \partial t_I \partial t_J$ equals an integrated two-point function, which in a unitary CFT it must have certain positivity properties.  These positivity properties lead to the conclusion that $F$ has a local maximum at $t_I= t_{I*}$.

To be more precise, at the SCFT fixed point, the correlation functions of the canonically normalized flavor currents and R-symmetry current take the form 
 \PUes{PUjjDef}{
   \langle j_I^\mu(x) j_J^\nu(0) \rangle  &= \frac{\tau_{IJ}}{16 \pi^2} 
   \left( \delta^{\mu\nu} \partial^2 - \partial^\mu \partial^\nu \right) \frac{1}{x^2}
    + \frac{i \kappa_{IJ}}{2 \pi} \epsilon^{\mu\nu\rho} \partial_\rho \delta^{(3)}(x) \,, \\
   \langle j_R^\mu(x) j_R^\nu(0) \rangle  &= \frac{\tau_{RR}}{16 \pi^2} 
   \left( \delta^{\mu\nu} \partial^2 - \partial^\mu \partial^\nu \right) \frac{1}{x^2}
    + \frac{i \kappa_{RR}}{2 \pi} \epsilon^{\mu\nu\rho} \partial_\rho \delta^{(3)}(x) \,, \\
  \langle j_I^\mu(x) j_R^\nu(0) \rangle  &=  
    \frac{i \kappa_{IR}}{2 \pi} \epsilon^{\mu\nu\rho} \partial_\rho \delta^{(3)}(x) \,,
 }
where $\tau_{IJ}$ and $\tau_{RR}$ are universal real constants that are positive by unitarity, while the contact terms proportional to the real coefficients $\kappa_{IJ}$, $\kappa_{IR}$ and $\kappa_{RR}$ can in general depend on the precise UV completion of the theory.  By relating the correlation functions of other operators in the flavor current and ${\cal R}_\mu$ multiplets to \eqref{PUjjDef}, one can show that   \cite{PUClosset:2012vg}
 \PUes{PUFExpansion}{
  -\log  Z(t_I) = - \log Z(t_{I*}) + i 2 \pi \kappa_{IR} (t_I - t_{I*})
   - \frac 12 \left(\frac{\pi^2}{2} \tau_{IJ} - 2 \pi i \kappa_{IJ}  \right) (t_I - t_{I*}) (t_J - t_{J*}) + \cdots \,.
 }
Recalling that we defined $F = - \log \abs{Z} = -\text{Re}\, \log Z$, we can infer from \eqref{PUFExpansion} that
 \PUes{PUFDers}{
  \frac{\partial F}{\partial t_I} \bigg|_{t = t_*} = 0 \,, \qquad
   \frac{\partial^2 F}{\partial t_I \partial t_J} \bigg|_{t = t_*} = - \frac{\pi^2}{2} \tau_{IJ} \,.
 }
Unitarity requires $\tau_{IJ} > 0$, and so \eqref{PUFDers} provides a proof of the $F$-maximization principle.

It is worth emphasizing that, as described above, the function $F(t_I)$ carries useful information about the SCFT even away from $t_I = t_{I*}$.  In particular, one can extract the two-point function coefficients $\tau_{IJ}$ from the second derivative of $F$ \cite{PUClosset:2012vg}:
 \PUes{PUGotTau}{
  \tau_{IJ} = -\frac{2}{\pi^2} \frac{\partial^2 F}{\partial t_I \partial t_J} \bigg|_{t = t_*} \,.
 }
Such a relation currently provides the only way of calculating the value of $\tau_{IJ}$ in strongly-coupled SCFTs and has been used, for instance, as a key input in the 3-d supersymmetric conformal bootstrap analysis of various ${\cal N} = 2$ SCFTs \cite{PUChester:2015qca,PUChester:2015lej,PUChester:2014fya,PUBobev:2015jxa,PUBeem:2016cbd}. 

Note that $F$-maximization implies that in the case where there are no accidental symmetries at the IR fixed point, the $F$-coefficient does decrease under supersymmetric RG flows triggered by superpotential deformations \cite{PUClosset:2012vg}.  Indeed, at the UV fixed point, where one should neglect the superpotential, there are more flavor symmetries that can mix with the R-symmetry, so $F$-maximization has to be performed over a larger set of trial R-charges than in the presence of the superpotential.  Consequently, $F_\text{UV} > F_\text{IR}$ in these examples.

\subsection{Examples}
\label{PULOCEXAMPLES}

\subsubsection{${\cal N} = 2$ super-Ising CFT and Wess-Zumino models}
\label{PUWESSZUMINO}

Perhaps the simplest example of an ${\cal N} = 2$ SCFT where one can use the methods described above to compute its $F$ coefficient is the critical ${\cal N} = 2$ super-Ising model.  It can be described in terms of a single chiral multiplet $\Phi = (Z, \chi, F)$ with a cubic superpotential interaction $W = g \Phi^3$, $g$ being a dimensionful coupling constant.  This non-conformal theory is believed to flow in the infrared to an ${\cal N} = 2$ SCFT---the ${\cal N} =2$ super-Ising CFT\@.  The superpotential does not preserve any flavor symmetries.  The only R-symmetry is that under which $\Phi$ has R-charge $r = 1/3$.  The $F$-coefficient of this theory can be read off from \eqref{PUGotZLoc} to be
 \PUes{PUFSuperIsing}{
  F_\text{${\cal N} = 2$ Ising} = -\ell(2/3) \approx 0.259 \,.
 }
This value is smaller than that of a free chiral multiplet $F_\text{chiral} \approx 0.347 $ (see \eqref{PUGotFN2}), which is the UV CFT fixed point of the RG flow $W = g \Phi^3$.

The ${\cal N} = 2$ super-Ising CFT is one of many Wess-Zumino models \cite{PUWess:1974tw} that are believed to flow to interacting SCFTs in the infrared.  (For a review, see~\cite{PUStrassler:2003qg}.)  For instance, one can construct supersymmetric generalization of the critical $O(N)$ vector model by considering $N+1$ chiral multiplets $\Phi_i = (Z_i, \chi_i, F_i)$ with the $O(N)$-invariant cubic superpotential
 \PUes{PUCubic}{
  W = g\, \Phi_{N+1} \sum_{i=1}^N \Phi_i^2 \,.
 }
These models have been studied in the $1/N$ expansion \cite{PUFerreira:1997he}, the $4-\epsilon$ expansion \cite{PUFerreira:1997hx,PUFerreira:1996az,PUJack:1999fa,PUJack:1998iy}, and more recently using supersymmetric localization \cite{PUnishioka2013rg, PUGiombi:2014xxa} and the conformal bootstrap \cite{PUChester:2015qca, PUChester:2015lej}.\footnote{These models provide a counterexample to a possible conjecture stating that the coefficient $c_T$ appearing in the two-point function of the canonically normalized stress tensor decreases along RG flow \cite{PUnishioka2013rg}.}  The RG flow triggered by \eqref{PUCubic} preserves a $U(1)_R$ symmetry as well as an $O(N) \times U(1)$ flavor symmetry.  Under the $O(N)$ symmetry, the $\Phi_i$ transform as a vector and $\Phi_{N+1}$ is a singlet, while under the flavor $U(1)$ $\Phi_i$ has charge $+1$ for $i =1, \ldots, N$ and charge $-2$ for $i = N+1$.  Since there is one Abelian flavor symmetry, there is a one-parameter family of R-charge assignments consistent with the $O(N)$-invariance of the theory and with the fact that the superpotential has R-charge two:
 \PUes{PURchargeAssignments}{
  r_i &= r \,, \qquad \text{for $i = 1, \ldots, N$} \,, \\
  r_{N+1}&= 2-2 r \,.
 }
Using \eqref{PUGotZLoc}, one finds
 \PUes{PUFWZ}{
  F = -N \ell(1-r) - \ell( 2r - 1) \,.
 }
The $F$-maximization principle states that one should maximize \eqref{PUFWZ} with respect to $r$ in order to find the value of $r$ attained at the SCFT fixed point.  Doing so, one obtains the values of $r$ given in Table~\ref{PURCharges}.  Since the multiplets $\Phi_i = (Z_i, \chi_i, F_i)$ are chiral, the values of $r$ given in Table~\ref{PURCharges} also determine the scaling dimensions of $Z_i$ and $\chi_i$ to be $\Delta_{Z_i} = r_i$ and $\Delta_{\chi_i} = r_i + \frac 12$, respectively.
\begin{table}
\begin{center}
  \begin{tabular}{c|c|c|c|c|c|c|c|c|c|c}
   $N$ & $1$ & $2$ & $3$ & $4$ & $5$ & $6$ & $7$ & $8$ & $9$ & $10$  \\
 \hline\hline
 $r_i = r$ (for $i=1, \ldots, N$) & $.708$ & $.667$ & $.632$ & $.605$ & $.586$ & $.572$ & $.562$ & $.554$ & $.548$ & $.543$ \\
 \hline
 $r_{N+1} = 2 - 2r$ & $.584$ & $.667$ & $.737$ & $.790$ & $.828$ & $.856$ & $.876$ & $.892$ & $.904$ & $.914$ 
  \end{tabular}
\end{center}
\caption{The superconformal R-charges $r_i$ at the IR fixed point of \eqref{PUCubic}. }
\label{PURCharges}
\end{table}%

In the case $N=2$, one can make the redefinitions $X = \Phi_3$, $Y = \Phi_1+i \Phi_2$, and $Z = \Phi_1 - i \Phi_2$ and rewrite the superpotential \eqref{PUCubic} as $W = g\, X Y Z$.  This theory is the ``$XYZ$ model.''  It is invariant under permuting $X$, $Y$, and $Z$, and consequently at the IR fixed point one expect the R-charges of $X$, $Y$, and $Z$ to be equal.  Since these charges must add up to two, we must have $r_X = r_Y = r_Z = 2/3$.  Indeed, one can check that when $N=2$, \eqref{PUFWZ} is maximized when $r = 2/3$.

Wess-Zumino models of the same type as above also provide an example that emphasizes the limitation of the $F$-maximization principle of not incorporating accidental symmetries.  When using $F$-maximization, one assumes that the RG flow ends at an SCFT where no accidental symmetries are present.  This assumption may of course be wrong.  For instance, one can show that the assumption of no accidental symmetries  is indeed incorrect in the following generalization of \eqref{PUCubic}:
 \PUes{PUCubicNo}{
  W = g_1\, \Phi_{N+1} \sum_{i=1}^N \Phi_i^2 + g_2\, \Phi_{N+1}^3 \,,
 }
where $g_1$ and $g_2$ are coupling constants.  The RG flow triggered by \eqref{PUCubicNo} preserves an $O(N) \times \BZ_3$ flavor symmetry and a unique $U(1)_R$ symmetry under which all the $\Phi_i$ have R-charge $2/3$.  It is tempting to assume that the IR limit of \eqref{PUCubicNo} is a unitary SCFT where all the $\Phi_i$ have R-charge $2/3$, but this assumption was recently proven to be incorrect if $N>2$ using the conformal bootstrap \cite{PUChester:2015qca}.  Based on arguments coming from the $4-\epsilon$ expansion, what is believed to happen in the model \eqref{PUCubicNo} when $N>2$ is that the coupling $g_2$ flows to zero in the IR, the flavor symmetry thus being enhanced to $O(N) \times U(1)$.  The IR fixed point of \eqref{PUCubicNo} is then believed to be the same as that of \eqref{PUCubic}.

\subsubsection{${\cal N} = 2$ SQED and a test of dualities}
\label{PUSQED}

As another example, one can consider ${\cal N} = 2$ supersymmetric quantum electrodynamics with $N$ pairs of conjugate flavors that we denote by $\Phi_i$ (of gauge charge $+1$) and $\tilde \Phi_i$ (of gauge charge $-1$), with $i=1, \ldots, N$, and vanishing superpotential.  One can also add a Chern-Simons term with level $k$ for the $U(1)$ vector multiplet.   This theory has the following flavor symmetries:  a $U(1)$ flavor symmetry under which both $\Phi_i$ and $\tilde \Phi_i$ have the same charge;    an $SU(N)$ symmetry under which the $\Phi_i$ transform as a fundamental and $\tilde \Phi_i$ as an anti-fundamental;  a topological $U(1)$ symmetry generated by $*F$; as well as a charge conjugation symmetry that flips the sign of the fields in the vector multiplet and interchanges $\Phi_i$ with $\tilde \Phi_i$. If we want to preserve the $SU(N)$ symmetry and the charge conjugation symmetry, then the R-symmetry can only mix with the flavor $U(1)$ under which $\Phi_i$ and $\tilde \Phi_i$ have the same charge.  Thus, one can consider a family of R-charge assignments
 \PUes{PURChargeSQED}{
  r_{\Phi_i} = r_{\tilde \Phi_i} = r
 }
parameterized by $r$.   The $S^3$ partition function is\footnote{We set $r_\text{top} = 0$, thus assigning the monopole operators with topological charge $+ q_\text{top}$ and $- q_\text{top}$ equal R-charges.  Such an assignment is consistent with the charge conjugation symmetry.  When $k \neq 0$, this assignment can also be thought of as fixing the flat direction mentioned in Footnote~\ref{PUFlat}.}
 \PUes{PUZSQED}{
  Z(r) = \int_{-\infty}^\infty d\sigma e^{i \pi k \sigma^2} e^{ N \left( \ell(1 - r + i \sigma) +\ell(1 - r - i \sigma)\right) } \,.
 }
It is straightforward to calculate numerically this integral and maximize $F = - \log \abs{Z}$ with respect to $r$.  One can also develop an analytical approximation in the regime where $N$ and $k$ are both taken to be large, with the ratio $\kappa = 2 k / (N \pi)$ fixed.  One finds the value of $r$ that maximizes $F$ to be \cite{PUKlebanov:2011td}
 \PUes{PUGotRSQED}{
  r = \frac 12  - \frac{2}{\pi^2 ( 1 + \kappa^2)} \frac 1N - \frac{2 \left[ \pi^2 - 12 + \kappa^2 (4 - 2 \pi^2) + \pi^2 \kappa^4 \right]}{\pi^4 (1 + \kappa^2)^3} \frac{1}{N^2} + O(N^{-3}) \,.
 }
The corresponding $F$-coefficient is 
 \PUes{PUGotFSQED}{
  F = N \log 2 + \frac 12 \log \left( \frac{N \pi}{2} \sqrt{1 + \kappa^2} \right) + \left( \frac{\kappa^2 - 1}{4 (1 + \kappa^2)^2} 
   + \frac{2}{\pi^2 (1 + \kappa^2)^2} \right) \frac 1N + O(N^{-2}) \,.
 }
It can be checked that the analytical approximation \eqref{PUGotRSQED}--\eqref{PUGotFSQED} matches quite well the numerical results even at fairly small values of $N$.  It also matches the large $N$ expansion in \eqref{PUFComplicated}, if one identifies $N_b = N_f = 2N$.  See \cite{PUKlebanov:2011td} for more details.

An interesting particular case is SQED with only one pair of conjugate chiral multiplets of unit gauge charge and no Chern-Simons interactions, namely $k=0$ and $N = 1$, where it can be checked numerically that the value of $r$ that maximizes $F$ is $r = 1/3$.  Indeed, in this case the $S^3$ partition function can be written as \cite{PUJafferis:2010un}
 \PUes{PUPartFunctionSQEDOne}{
  Z(r) =  \int_{-\infty}^\infty d\sigma\, e^{  \ell(1 - r + i \sigma) +\ell(1 - r - i \sigma) }
   = e^{2 \ell(r) + \ell(1-2 r) } \,,
 }
where the last equality can be checked numerically, for instance.  This is nothing but the $S^3$ partition function of the $XYZ$ model (see the discussion following \eqref{PUGotZLoc}) with R-charge assignments $r_X = 2r$, $r_Y = 1-r$, $r_Z = 1-r$.  Indeed, the SQED with $1$ pair of conjugate chirals is known to be dual to the $XYZ$ model \cite{PUAharony:1997bx}.  That $F$ is maximized when $r = 1/3$ is consistent with the fact that in the $XYZ$ model $F$ is maximized for the symmetric R-charge assignment $r_X = r_Y = r_Z = 2/3$.  The expression \eqref{PUPartFunctionSQEDOne} is not just a check of the duality between the $XYZ$ model and SQED, but it also provides some insight into how the duality works.  In particular, the chiral field $X$ is dual to an operator of R-charge $2r$ (this is $\tilde Q Q$), while $Y$ and $Z$ are dual to operators of R-charge $1-r$ (these are monopole operators).  Other tests of dualities using the $S^3$ partition function were performed, for example, in \cite{PUWillett:2011gp,PUKapustin:2010xq,PUKapustin:2010mh,PUAmariti:2011uw,PUSafdi:2012re,PUKapustin:2011gh,PUJafferis:2011ns,PUKapustin:2011vz,PUMorita:2011cs,PUBenini:2011mf}.

\subsubsection{Examples in holography}

The ${\cal N} = 2$ SCFTs with holographic duals provide a richer set of examples in which one can calculate $F$ via supersymmetric localization and compare it to the supergravity expectation.  For instance, there are many SCFTs that are conjectured to be dual to M-theory backgrounds of the form $AdS_4 \times Y_7$, where $Y_7$ is a Sasaki-Einstein space, realized by placing $N$ coincident M2-branes at tip of the Calabi-Yau cone over $Y_7$.  In these instances, supergravity predicts that the $F$-coefficient is given by \eqref{PUFVolY}.  There have been many field theory computations of $F$ in ${\cal N} = 2$ SCFTs that match this supergravity result.  See, for example, \cite{PUJafferis:2011zi,PUGabella:2011sg,PUMartelli:2011qj,PUGang:2011jj,PUAmariti:2011uw,PUGulotta:2011aa,PUKim:2012vza}.

Moving away from SCFTs, one may wonder whether it is possible to calculate the $S^3$ in supergravity and reproduce from a holographic computation the entire function of the trial R-charges, even before performing $F$-maximization in the field theory.  This question was studied in \cite{PUFreedman:2013ryh} in the context of ABJM theory \cite{PUAharony:2008ug}.   ABJM theory is a $U(N)_k \times U(N)_{-k}$ Chern-Simons-matter theory that in general preserves ${\cal N} = 6$ supersymmetry that is believed to be enhanced to ${\cal N} = 8$ when $k=1$ or $2$.  In ${\cal N} = 2$ notation, the field content of ABJM theory consists of two $U(N)$ vector multiplets with Chern-Simons interactions $k$ and $-k$, with matter content consisting of two chiral multiplets ${\cal Z}_i$, $i=1, 2$ transforming as a bifundamental of $U(N) \times U(N)$ and two chiral multiplets ${\cal W}_i$, $i=1, 2$, transforming in the conjugate representation.  The superpotential is of the form
 \PUes{PUWABJM}{
  W \propto \epsilon^{ik} \epsilon^{jl} \tr\left[ {\cal W}_i {\cal Z}_j {\cal W}_k {\cal Z}_l \right] \,,
 }
with a precise coefficient fixed by the extended supersymmetry of the theory.  The extended supersymmetry also fixes the R-charges of the chiral operators to $r_{{\cal W}_i} = r_{{\cal Z}_i} = 1/2$, so there is no need to perform $F$-maximization in this case.  However, one can nevertheless consider a 3-parameter family of trial R-charges given by arbitrary $r_{{\cal W}_i}$ and $r_{{\cal Z}_i}$ with the constraint 
 \PUes{PUTrialABJM}{
  r_{{\cal W}_1} + r_{{\cal W}_2} + r_{{\cal Z}_1} + r_{{\cal Z}_2} = 2
 }
that ensures that the superpotential has R-charge $2$.\footnote{We may $r_\text{top}^a = 0$ as a choice in order to fix the flat directions mentioned in Footnote~\ref{PUFlat}.  These parameters were included in the analysis performed in \cite{PUJafferis:2011zi}.}  This R-charge assignment preserves only ${\cal N} = 2$ supersymmetry.  Using \eqref{PUGotZLoc} and the matrix model technique developed in \cite{PUHerzog:2010hf} one can show that at large $N$ the $S^3$ free energy takes the form \cite{PUJafferis:2011zi}
 \PUes{PUABJMZTrial}{
  F = \frac{\sqrt{2}\pi k^{1/2} N^{3/2}}{3} 4 \sqrt{r_{{\cal W}_1} r_{{\cal W}_2} r_{{\cal Z}_1} r_{{\cal Z}_2}} + O(N^{1/2})  \,.
 } 
This expression agrees with \eqref{PUFreeABJM} when $r_{{\cal W}_i} = r_{{\cal Z}_i} = 1/2$.

In the case $k=1$, the 3-parameter R-charge deformations mentioned above are dual to holographic RG flows that asymptote to $\BH^4 \times S^7$ in the UV\@.  These flows were constructed in \cite{PUFreedman:2013ryh} in a 4-d model that can be uplifted to a background of 11-d supergravity.  This model involves Einstein gravity coupled to three complex scalar fields, each of which corresponds to one of the three parameters in the family of R-charge assignments. Upon a careful use of holographic renormalization and  supersymmetry, Ref.~\cite{PUFreedman:2013ryh} obtained a perfect match of the 4-d on-shell supergravity action with \eqref{PUABJMZTrial}.\footnote{See also~\cite{PUMartelli:2011fu,PUMartelli:2011fw,PUMartelli:2012sz,PUMartelli:2013aqa,PUFarquet:2014kma,PUFarquet:2013cwa,PUFarquet:2014bda,PUBobev:2013cja,PUBobev:2016nua} for other constructions of supergravity backgrounds dual to deformations of supersymmetric field theories on curved manifolds.}

\section{Conclusion}
\label{PUDISCUSSION}

In this contribution I reviewed some of the recent developments related to the $S^3$ free energy of various supersymmetric and non-supersymmetric CFTs in three dimensions, in particular the $F$-theorem and the $F$-maximization principle and some of their applications.  In Section~\ref{PUFCOEFFICIENTS} I have shown how the $F$-coefficient can be computed in various approximation schemes, and how these results are consistent with the $F$-theorem in several examples.  In Section~\ref{PUFMAXIMIZATION} I explained how $F$ can be computed exactly in SCFTs with ${\cal N} \geq 2$ supersymmetry, and as a byproduct how one can determine the R-charges (or scaling dimensions) of the various chiral operators of these SCFTs by maximizing the $F$ over a set of trial R-charges, both in general and in a few examples.

The irreversibility of the RG trajectories that is required by the $F$-theorem is a fundamental property of relativistic quantum field theory in three dimensions.  An interesting open problem remains to prove the $F$-theorem in a way that uses directly the properties of the $S^3$ partition function, without appealing to the notion of entanglement entropy.  Perhaps a related future direction would be to construct a function that interpolates between $F_\text{UV}$ and $F_\text{IR}$ monotonically along any RG trajectory and that is stationary at the UV and IR fixed points.  The proof of the $F$-theorem using entanglement entropy \cite{PUCasini:2012ei,PUCasini:2011kv} that I did not review here does provide a strictly monotonic interpolating function, namely the renormalized entanglement entropy proposed in \cite{PULiu:2012eea}, but this function may or may not be stationary at the UV and IR fixed points \cite{PUKlebanov:2012va,PULiu:2013una}.  Lastly, it would be interesting to investigate whether there exists an analog of the $F$-theorem in a larger odd number of spacetime dimensions.  For instance, in five dimensions there are a few examples of RG trajectories between pairs of CFTs that obey a conjectured $F$-theorem \cite{PUJafferis:2012iv,PUFei:2014yja}, but there is no general proof of such a result.

\section*{Acknowledgments}

I am particularly grateful to Shai Chester, Dan Freedman, Simone Giombi, Dan Gulotta, Chris Herzog, Luca Iliesiu, Daniel Jafferis, Igor Klebanov, Jaehoon Lee, Tatsuma Nishioka, Subir Sachdev, Ben Safdi, Grisha Tarnopolsky, Tibi Tesileanu, and Ran Yacoby for collaboration on the various projects mentioned or reviewed here.   This work was supported in part by the US NSF under Grant No.~PHY-1418069.

\documentfinish
\begin{thebibliography}{100}

\bibitem{ContributionSummary}
V.~Pestun and M.~Zabzine, eds., {\em Localization techniques in quantum field
  theory}, vol.~xx.
\newblock Journal of Physics A, 2016.
\newblock \href{http://arxiv.org/abs/1608.02952}{{\tt 1608.02952}}.
\newblock \url{https://arxiv.org/src/1608.02952/anc/LocQFT.pdf},
  \url{http://pestun.ihes.fr/pages/LocalizationReview/LocQFT.pdf}.

\bibitem{PUJackiw:2011vz}
R.~Jackiw and S.~Y. Pi, ``{Tutorial on Scale and Conformal Symmetries in
  Diverse Dimensions},''
  \href{http://dx.doi.org/10.1088/1751-8113/44/22/223001}{{\em J. Phys.} {\bf
  A44} (2011)  223001},
\href{http://arxiv.org/abs/1101.4886}{{\tt arXiv:1101.4886 [math-ph]}}.
%%CITATION = ARXIV:1101.4886;%%.

\bibitem{PUDymarsky:2013pqa}
A.~Dymarsky, Z.~Komargodski, A.~Schwimmer, and S.~Theisen, ``{On Scale and
  Conformal Invariance in Four Dimensions},''
\href{http://arxiv.org/abs/1309.2921}{{\tt arXiv:1309.2921 [hep-th]}}.
%%CITATION = ARXIV:1309.2921;%%.

\bibitem{PUDymarsky:2014zja}
A.~Dymarsky, K.~Farnsworth, Z.~Komargodski, M.~A. Luty, and V.~Prilepina,
  ``{Scale Invariance, Conformality, and Generalized Free Fields},''
\href{http://arxiv.org/abs/1402.6322}{{\tt arXiv:1402.6322 [hep-th]}}.
%%CITATION = ARXIV:1402.6322;%%.

\bibitem{PUZamolodchikov:1986gt}
A.~Zamolodchikov, ``{Irreversibility of the Flux of the Renormalization Group
  in a 2D Field Theory},''
{\em JETP Lett.} {\bf 43} (1986)  730--732.
%%CITATION = JTPLA,43,730;%%.

\bibitem{PUCardy:1988cwa}
J.~L. Cardy, ``{Is There a c Theorem in Four-Dimensions?},''
\href{http://dx.doi.org/10.1016/0370-2693(88)90054-8}{{\em Phys.Lett.} {\bf
  B215} (1988)  749--752}.
%%CITATION = PHLTA,B215,749;%%.

\bibitem{PUKomargodski:2011vj}
Z.~Komargodski and A.~Schwimmer, ``{On Renormalization Group Flows in Four
  Dimensions},'' \href{http://dx.doi.org/10.1007/JHEP12(2011)099}{{\em JHEP}
  {\bf 1112} (2011)  099},
\href{http://arxiv.org/abs/1107.3987}{{\tt arXiv:1107.3987 [hep-th]}}.
%%CITATION = ARXIV:1107.3987;%%.

\bibitem{PUAffleck:1992ng}
I.~Affleck and A.~W.~W. Ludwig, ``{Exact conformal-field-theory results on the
  multichannel Kondo effect: Single-fermion Green's function, self-energy, and
  resistivity},''
\href{http://dx.doi.org/10.1103/PhysRevB.48.7297}{{\em Phys. Rev.} {\bf B48}
  (1993) no.~10, 7297}.
%%CITATION = PHRVA,B48,7297;%%.

\bibitem{PUAffleck:1991tk}
I.~Affleck and A.~W.~W. Ludwig, ``{Universal noninteger 'ground state
  degeneracy' in critical quantum systems},''
\href{http://dx.doi.org/10.1103/PhysRevLett.67.161}{{\em Phys. Rev. Lett.} {\bf
  67} (1991)  161--164}.
%%CITATION = PRLTA,67,161;%%.

\bibitem{PUJafferis:2010un}
D.~L. Jafferis, ``{The Exact Superconformal R-Symmetry Extremizes $Z$},''
  \href{http://dx.doi.org/10.1007/JHEP05(2012)159}{{\em JHEP} {\bf 1205} (2012)
   159},
\href{http://arxiv.org/abs/1012.3210}{{\tt arXiv:1012.3210 [hep-th]}}.
%%CITATION = ARXIV:1012.3210;%%.

\bibitem{PUJafferis:2011zi}
D.~L. Jafferis, I.~R. Klebanov, S.~S. Pufu, and B.~R. Safdi, ``{Towards the
  F-Theorem: ${\cal N}=2$ Field Theories on the Three-Sphere},''
  \href{http://dx.doi.org/10.1007/JHEP06(2011)102}{{\em JHEP} {\bf 1106} (2011)
   102},
\href{http://arxiv.org/abs/1103.1181}{{\tt arXiv:1103.1181 [hep-th]}}.
%%CITATION = ARXIV:1103.1181;%%.

\bibitem{PUClosset:2012vg}
C.~Closset, T.~T. Dumitrescu, G.~Festuccia, Z.~Komargodski, and N.~Seiberg,
  ``{Contact Terms, Unitarity, and F-Maximization in Three-Dimensional
  Superconformal Theories},''
  \href{http://dx.doi.org/10.1007/JHEP10(2012)053}{{\em JHEP} {\bf 1210} (2012)
   053},
\href{http://arxiv.org/abs/1205.4142}{{\tt arXiv:1205.4142 [hep-th]}}.
%%CITATION = ARXIV:1205.4142;%%.

\bibitem{PUIntriligator:2003jj}
K.~A. Intriligator and B.~Wecht, ``{The Exact superconformal R symmetry
  maximizes a},'' \href{http://dx.doi.org/10.1016/S0550-3213(03)00459-0}{{\em
  Nucl.Phys.} {\bf B667} (2003)  183--200},
\href{http://arxiv.org/abs/hep-th/0304128}{{\tt arXiv:hep-th/0304128
  [hep-th]}}.
%%CITATION = HEP-TH/0304128;%%.

\bibitem{PUGulotta:2011si}
D.~R. Gulotta, C.~P. Herzog, and S.~S. Pufu, ``{From Necklace Quivers to the
  F-theorem, Operator Counting, and T(U(N))},''
  \href{http://dx.doi.org/10.1007/JHEP12(2011)077}{{\em JHEP} {\bf 1112} (2011)
   077},
\href{http://arxiv.org/abs/1105.2817}{{\tt arXiv:1105.2817 [hep-th]}}.
%%CITATION = ARXIV:1105.2817;%%.

\bibitem{PUKlebanov:2011gs}
I.~R. Klebanov, S.~S. Pufu, and B.~R. Safdi, ``{F-Theorem without
  Supersymmetry},'' \href{http://dx.doi.org/10.1007/JHEP10(2011)038}{{\em JHEP}
  {\bf 1110} (2011)  038},
\href{http://arxiv.org/abs/1105.4598}{{\tt arXiv:1105.4598 [hep-th]}}.
%%CITATION = ARXIV:1105.4598;%%.

\bibitem{PUKlebanov:2011td}
I.~R. Klebanov, S.~S. Pufu, S.~Sachdev, and B.~R. Safdi, ``{Entanglement
  Entropy of 3-d Conformal Gauge Theories with Many Flavors},''
  \href{http://dx.doi.org/10.1007/JHEP05(2012)036}{{\em JHEP} {\bf 1205} (2012)
   036},
\href{http://arxiv.org/abs/1112.5342}{{\tt arXiv:1112.5342 [hep-th]}}.
%%CITATION = ARXIV:1112.5342;%%.

\bibitem{PUMyers:2010xs}
R.~C. Myers and A.~Sinha, ``{Seeing a c-theorem with holography},''
  \href{http://dx.doi.org/10.1103/PhysRevD.82.046006}{{\em Phys.Rev.} {\bf D82}
  (2010)  046006},
\href{http://arxiv.org/abs/1006.1263}{{\tt arXiv:1006.1263 [hep-th]}}.
%%CITATION = ARXIV:1006.1263;%%.

\bibitem{PUMyers:2010tj}
R.~C. Myers and A.~Sinha, ``{Holographic c-theorems in arbitrary dimensions},''
  \href{http://dx.doi.org/10.1007/JHEP01(2011)125}{{\em JHEP} {\bf 1101} (2011)
   125},
\href{http://arxiv.org/abs/1011.5819}{{\tt arXiv:1011.5819 [hep-th]}}.
%%CITATION = ARXIV:1011.5819;%%.

\bibitem{PUCasini:2011kv}
H.~Casini, M.~Huerta, and R.~C. Myers, ``{Towards a derivation of holographic
  entanglement entropy},''
  \href{http://dx.doi.org/10.1007/JHEP05(2011)036}{{\em JHEP} {\bf 1105} (2011)
   036},
\href{http://arxiv.org/abs/1102.0440}{{\tt arXiv:1102.0440 [hep-th]}}.
%%CITATION = ARXIV:1102.0440;%%.

\bibitem{PULiu:2012eea}
H.~Liu and M.~Mezei, ``{A Refinement of entanglement entropy and the number of
  degrees of freedom},'' \href{http://dx.doi.org/10.1007/JHEP04(2013)162}{{\em
  JHEP} {\bf 1304} (2013)  162},
\href{http://arxiv.org/abs/1202.2070}{{\tt arXiv:1202.2070 [hep-th]}}.
%%CITATION = ARXIV:1202.2070;%%.

\bibitem{PUCasini:2012ei}
H.~Casini and M.~Huerta, ``{On the RG running of the entanglement entropy of a
  circle},'' \href{http://dx.doi.org/10.1103/PhysRevD.85.125016}{{\em
  Phys.Rev.} {\bf D85} (2012)  125016},
\href{http://arxiv.org/abs/1202.5650}{{\tt arXiv:1202.5650 [hep-th]}}.
%%CITATION = ARXIV:1202.5650;%%.

\bibitem{PUCasini:2006es}
H.~Casini and M.~Huerta, ``{A c-theorem for the entanglement entropy},''
  \href{http://dx.doi.org/10.1088/1751-8113/40/25/S57}{{\em J. Phys.} {\bf A40}
  (2007)  7031--7036},
\href{http://arxiv.org/abs/cond-mat/0610375}{{\tt arXiv:cond-mat/0610375
  [cond-mat]}}.
%%CITATION = COND-MAT/0610375;%%.

\bibitem{PUQuineChoi}
J.~R. Quine and J.~Choi, ``{Zeta regularized products and functional
  determinants on spheres},'' {\em Rocky Mountain J. Math.} {\bf 26} (1996)
  no.~2, 719--729.

\bibitem{PUKumagai}
H.~Kumagai, ``{The determinant of the Laplacian on the $n$-sphere},'' {\em Acta
  Arith.} {\bf 91} (1999) no.~3, 199--208.

\bibitem{PUMarino:2011nm}
M.~Marino, ``{Lectures on localization and matrix models in supersymmetric
  Chern-Simons-matter theories},''
  \href{http://dx.doi.org/10.1088/1751-8113/44/46/463001}{{\em J.Phys.A} {\bf
  A44} (2011)  463001}, \href{http://arxiv.org/abs/1104.0783}{{\tt
  arXiv:1104.0783 [hep-th]}}.

\bibitem{PUWitten:1988hf}
E.~Witten, ``{Quantum Field Theory and the Jones Polynomial},''
\href{http://dx.doi.org/10.1007/BF01217730}{{\em Commun. Math. Phys.} {\bf 121}
  (1989)  351--399}.
%%CITATION = CMPHA,121,351;%%.

\bibitem{PUAgon:2013iva}
C.~A. Agon, M.~Headrick, D.~L. Jafferis, and S.~Kasko, ``{Disk entanglement
  entropy for a Maxwell field},''
  \href{http://dx.doi.org/10.1103/PhysRevD.89.025018}{{\em Phys. Rev.} {\bf
  D89} (2014) no.~2, 025018},
\href{http://arxiv.org/abs/1310.4886}{{\tt arXiv:1310.4886 [hep-th]}}.
%%CITATION = ARXIV:1310.4886;%%.

\bibitem{PUGiombi:2013yva}
S.~Giombi, I.~R. Klebanov, S.~S. Pufu, B.~R. Safdi, and G.~Tarnopolsky, ``{AdS
  Description of Induced Higher-Spin Gauge Theory},''
  \href{http://dx.doi.org/10.1007/JHEP10(2013)016}{{\em JHEP} {\bf 1310} (2013)
   016},
\href{http://arxiv.org/abs/1306.5242}{{\tt arXiv:1306.5242 [hep-th]}}.
%%CITATION = ARXIV:1306.5242;%%.

\bibitem{PUWilson:1971dc}
K.~G. Wilson and M.~E. Fisher, ``{Critical exponents in 3.99 dimensions},''
\href{http://dx.doi.org/10.1103/PhysRevLett.28.240}{{\em Phys. Rev. Lett.} {\bf
  28} (1972)  240--243}.
%%CITATION = PRLTA,28,240;%%.

\bibitem{PUWilson:1973jj}
K.~G. Wilson and J.~B. Kogut, ``{The Renormalization group and the epsilon
  expansion},''
\href{http://dx.doi.org/10.1016/0370-1573(74)90023-4}{{\em Phys. Rept.} {\bf
  12} (1974)  75--200}.
%%CITATION = PRPLC,12,75;%%.

\bibitem{PUGiombi:2014xxa}
S.~Giombi and I.~R. Klebanov, ``{Interpolating between $a$ and $F$},''
  \href{http://dx.doi.org/10.1007/JHEP03(2015)117}{{\em JHEP} {\bf 03} (2015)
  117},
\href{http://arxiv.org/abs/1409.1937}{{\tt arXiv:1409.1937 [hep-th]}}.
%%CITATION = ARXIV:1409.1937;%%.

\bibitem{PUFei:2015oha}
L.~Fei, S.~Giombi, I.~R. Klebanov, and G.~Tarnopolsky, ``{Generalized
  $F$-Theorem and the $\epsilon$ Expansion},''
\href{http://arxiv.org/abs/1507.01960}{{\tt arXiv:1507.01960 [hep-th]}}.
%%CITATION = ARXIV:1507.01960;%%.

\bibitem{PUGiombi:2015haa}
S.~Giombi, I.~R. Klebanov, and G.~Tarnopolsky, ``{Conformal QED$_d$,
  $F$-Theorem and the $\epsilon$ Expansion},''
\href{http://arxiv.org/abs/1508.06354}{{\tt arXiv:1508.06354 [hep-th]}}.
%%CITATION = ARXIV:1508.06354;%%.

\bibitem{PUMinahan:2015any}
J.~A. Minahan, ``{Localizing gauge theories on S$^{d}$},''
  \href{http://dx.doi.org/10.1007/JHEP04(2016)152}{{\em JHEP} {\bf 04} (2016)
  152},
\href{http://arxiv.org/abs/1512.06924}{{\tt arXiv:1512.06924 [hep-th]}}.
%%CITATION = ARXIV:1512.06924;%%.

\bibitem{PUHerzog:2010hf}
C.~P. Herzog, I.~R. Klebanov, S.~S. Pufu, and T.~Tesileanu, ``{Multi-Matrix
  Models and Tri-Sasaki Einstein Spaces},''
  \href{http://dx.doi.org/10.1103/PhysRevD.83.046001}{{\em Phys. Rev.} {\bf
  D83} (2011)  046001},
\href{http://arxiv.org/abs/1011.5487}{{\tt arXiv:1011.5487 [hep-th]}}.
%%CITATION = ARXIV:1011.5487;%%.

\bibitem{PUKlebanov:1996un}
I.~R. Klebanov and A.~A. Tseytlin, ``{Entropy of near extremal black
  p-branes},'' \href{http://dx.doi.org/10.1016/0550-3213(96)00295-7}{{\em Nucl.
  Phys.} {\bf B475} (1996)  164--178},
\href{http://arxiv.org/abs/hep-th/9604089}{{\tt arXiv:hep-th/9604089
  [hep-th]}}.
%%CITATION = HEP-TH/9604089;%%.

\bibitem{PUWitten:1988ze}
E.~Witten, ``{Topological Quantum Field Theory},''
\href{http://dx.doi.org/10.1007/BF01223371}{{\em Commun. Math. Phys.} {\bf 117}
  (1988)  353}.
%%CITATION = CMPHA,117,353;%%.

\bibitem{PUPestun:2007rz}
V.~Pestun, ``{Localization of gauge theory on a four-sphere and supersymmetric
  Wilson loops},'' \href{http://dx.doi.org/10.1007/s00220-012-1485-0}{{\em
  Commun. Math. Phys.} {\bf 313} (2012)  71--129},
\href{http://arxiv.org/abs/0712.2824}{{\tt arXiv:0712.2824 [hep-th]}}.
%%CITATION = ARXIV:0712.2824;%%.

\bibitem{PUKapustin:2009kz}
A.~Kapustin, B.~Willett, and I.~Yaakov, ``{Exact Results for Wilson Loops in
  Superconformal Chern-Simons Theories with Matter},''
  \href{http://dx.doi.org/10.1007/JHEP03(2010)089}{{\em JHEP} {\bf 03} (2010)
  089},
\href{http://arxiv.org/abs/0909.4559}{{\tt arXiv:0909.4559 [hep-th]}}.
%%CITATION = ARXIV:0909.4559;%%.

\bibitem{PUSchwarz:2004yj}
J.~H. Schwarz, ``{Superconformal Chern-Simons theories},''
  \href{http://dx.doi.org/10.1088/1126-6708/2004/11/078}{{\em JHEP} {\bf 11}
  (2004)  078},
\href{http://arxiv.org/abs/hep-th/0411077}{{\tt arXiv:hep-th/0411077
  [hep-th]}}.
%%CITATION = HEP-TH/0411077;%%.

\bibitem{PUGaiotto:2007qi}
D.~Gaiotto and X.~Yin, ``{Notes on superconformal Chern-Simons-Matter
  theories},'' \href{http://dx.doi.org/10.1088/1126-6708/2007/08/056}{{\em
  JHEP} {\bf 08} (2007)  056},
\href{http://arxiv.org/abs/0704.3740}{{\tt arXiv:0704.3740 [hep-th]}}.
%%CITATION = ARXIV:0704.3740;%%.

\bibitem{ContributionWI}
B.~Willett, ``Localization on three-dimensional manifolds,'' {\em Journal of
  Physics A} {\bf xx} (2016)  000, \href{http://arxiv.org/abs/1608.02958}{{\tt
  1608.02958}}.

\bibitem{PUIntriligator:1996ex}
K.~A. Intriligator and N.~Seiberg, ``{Mirror symmetry in three-dimensional
  gauge theories},'' \href{http://dx.doi.org/10.1016/0370-2693(96)01088-X}{{\em
  Phys. Lett.} {\bf B387} (1996)  513--519},
\href{http://arxiv.org/abs/hep-th/9607207}{{\tt arXiv:hep-th/9607207
  [hep-th]}}.
%%CITATION = HEP-TH/9607207;%%.

\bibitem{PUDrukker:2010nc}
N.~Drukker, M.~Marino, and P.~Putrov, ``{From weak to strong coupling in ABJM
  theory},'' \href{http://dx.doi.org/10.1007/s00220-011-1253-6}{{\em Commun.
  Math. Phys.} {\bf 306} (2011)  511--563},
\href{http://arxiv.org/abs/1007.3837}{{\tt arXiv:1007.3837 [hep-th]}}.
%%CITATION = ARXIV:1007.3837;%%.

\bibitem{PUSantamaria:2010dm}
R.~C. Santamaria, M.~Marino, and P.~Putrov, ``{Unquenched flavor and tropical
  geometry in strongly coupled Chern-Simons-matter theories},''
  \href{http://dx.doi.org/10.1007/JHEP10(2011)139}{{\em JHEP} {\bf 10} (2011)
  139},
\href{http://arxiv.org/abs/1011.6281}{{\tt arXiv:1011.6281 [hep-th]}}.
%%CITATION = ARXIV:1011.6281;%%.

\bibitem{PUMarino:2011eh}
M.~Marino and P.~Putrov, ``{ABJM theory as a Fermi gas},''
  \href{http://dx.doi.org/10.1088/1742-5468/2012/03/P03001}{{\em J. Stat.
  Mech.} {\bf 1203} (2012)  P03001},
\href{http://arxiv.org/abs/1110.4066}{{\tt arXiv:1110.4066 [hep-th]}}.
%%CITATION = ARXIV:1110.4066;%%.

\bibitem{PUAharony:2008ug}
O.~Aharony, O.~Bergman, D.~L. Jafferis, and J.~Maldacena, ``{${\cal N}=6$
  superconformal Chern-Simons-matter theories, M2-branes and their gravity
  duals},'' \href{http://dx.doi.org/10.1088/1126-6708/2008/10/091}{{\em JHEP}
  {\bf 10} (2008)  091},
\href{http://arxiv.org/abs/0806.1218}{{\tt arXiv:0806.1218 [hep-th]}}.
%%CITATION = ARXIV:0806.1218;%%.

\bibitem{ContributionMA}
M.~Mari{\~n}o, ``Localization at large $N$ in Chern-Simons-matter theories,''
  {\em Journal of Physics A} {\bf xx} (2016)  000,
  \href{http://arxiv.org/abs/1608.02959}{{\tt 1608.02959}}.

\bibitem{PUMartelli:2011qj}
D.~Martelli and J.~Sparks, ``{The large N limit of quiver matrix models and
  Sasaki-Einstein manifolds},''
  \href{http://dx.doi.org/10.1103/PhysRevD.84.046008}{{\em Phys. Rev.} {\bf
  D84} (2011)  046008},
\href{http://arxiv.org/abs/1102.5289}{{\tt arXiv:1102.5289 [hep-th]}}.
%%CITATION = ARXIV:1102.5289;%%.

\bibitem{PUCheon:2011vi}
S.~Cheon, H.~Kim, and N.~Kim, ``{Calculating the partition function of ${\cal
  N}=2$ Gauge theories on $S^3$ and AdS/CFT correspondence},''
  \href{http://dx.doi.org/10.1007/JHEP05(2011)134}{{\em JHEP} {\bf 05} (2011)
  134},
\href{http://arxiv.org/abs/1102.5565}{{\tt arXiv:1102.5565 [hep-th]}}.
%%CITATION = ARXIV:1102.5565;%%.

\bibitem{PUGabella:2011sg}
M.~Gabella, D.~Martelli, A.~Passias, and J.~Sparks, ``{The free energy of
  ${\cal N}=2$ supersymmetric AdS$_4$ solutions of M-theory},''
  \href{http://dx.doi.org/10.1007/JHEP10(2011)039}{{\em JHEP} {\bf 10} (2011)
  039},
\href{http://arxiv.org/abs/1107.5035}{{\tt arXiv:1107.5035 [hep-th]}}.
%%CITATION = ARXIV:1107.5035;%%.

\bibitem{PUGulotta:2011vp}
D.~R. Gulotta, J.~P. Ang, and C.~P. Herzog, ``{Matrix Models for Supersymmetric
  Chern-Simons Theories with an ADE Classification},''
  \href{http://dx.doi.org/10.1007/JHEP01(2012)132}{{\em JHEP} {\bf 01} (2012)
  132},
\href{http://arxiv.org/abs/1111.1744}{{\tt arXiv:1111.1744 [hep-th]}}.
%%CITATION = ARXIV:1111.1744;%%.

\bibitem{PUGulotta:2012yd}
D.~R. Gulotta, C.~P. Herzog, and T.~Nishioka, ``{The ABCDEF's of Matrix Models
  for Supersymmetric Chern-Simons Theories},''
  \href{http://dx.doi.org/10.1007/JHEP04(2012)138}{{\em JHEP} {\bf 04} (2012)
  138},
\href{http://arxiv.org/abs/1201.6360}{{\tt arXiv:1201.6360 [hep-th]}}.
%%CITATION = ARXIV:1201.6360;%%.

\bibitem{PUCrichigno:2012sk}
P.~M. Crichigno, C.~P. Herzog, and D.~Jain, ``{Free Energy of $D_n$ Quiver
  Chern-Simons Theories},''
  \href{http://dx.doi.org/10.1007/JHEP03(2013)039}{{\em JHEP} {\bf 03} (2013)
  039},
\href{http://arxiv.org/abs/1211.1388}{{\tt arXiv:1211.1388 [hep-th]}}.
%%CITATION = ARXIV:1211.1388;%%.

\bibitem{PUAmariti:2012tj}
A.~Amariti and S.~Franco, ``{Free Energy vs Sasaki-Einstein Volume for Infinite
  Families of M2-Brane Theories},''
  \href{http://dx.doi.org/10.1007/JHEP09(2012)034}{{\em JHEP} {\bf 09} (2012)
  034},
\href{http://arxiv.org/abs/1204.6040}{{\tt arXiv:1204.6040 [hep-th]}}.
%%CITATION = ARXIV:1204.6040;%%.

\bibitem{PUAssel:2012cp}
B.~Assel, J.~Estes, and M.~Yamazaki, ``{Large N Free Energy of 3d N=4 SCFTs and
  $AdS_4/CFT_3$},'' \href{http://dx.doi.org/10.1007/JHEP09(2012)074}{{\em JHEP}
  {\bf 09} (2012)  074},
\href{http://arxiv.org/abs/1206.2920}{{\tt arXiv:1206.2920 [hep-th]}}.
%%CITATION = ARXIV:1206.2920;%%.

\bibitem{PUGuarino:2015jca}
A.~Guarino, D.~L. Jafferis, and O.~Varela, ``{String Theory Origin of Dyonic
  ${\cal N}=8$ Supergravity and Its Chern-Simons Duals},''
  \href{http://dx.doi.org/10.1103/PhysRevLett.115.091601}{{\em Phys. Rev.
  Lett.} {\bf 115} (2015) no.~9, 091601},
\href{http://arxiv.org/abs/1504.08009}{{\tt arXiv:1504.08009 [hep-th]}}.
%%CITATION = ARXIV:1504.08009;%%.

\bibitem{PUHama:2010av}
N.~Hama, K.~Hosomichi, and S.~Lee, ``{Notes on SUSY Gauge Theories on
  Three-Sphere},'' \href{http://dx.doi.org/10.1007/JHEP03(2011)127}{{\em JHEP}
  {\bf 03} (2011)  127},
\href{http://arxiv.org/abs/1012.3512}{{\tt arXiv:1012.3512 [hep-th]}}.
%%CITATION = ARXIV:1012.3512;%%.

\bibitem{PUFestuccia:2011ws}
G.~Festuccia and N.~Seiberg, ``{Rigid Supersymmetric Theories in Curved
  Superspace},'' \href{http://dx.doi.org/10.1007/JHEP06(2011)114}{{\em JHEP}
  {\bf 06} (2011)  114},
\href{http://arxiv.org/abs/1105.0689}{{\tt arXiv:1105.0689 [hep-th]}}.
%%CITATION = ARXIV:1105.0689;%%.

\bibitem{PUClosset:2012vp}
C.~Closset, T.~T. Dumitrescu, G.~Festuccia, Z.~Komargodski, and N.~Seiberg,
  ``{Comments on Chern-Simons Contact Terms in Three Dimensions},''
  \href{http://dx.doi.org/10.1007/JHEP09(2012)091}{{\em JHEP} {\bf 09} (2012)
  091},
\href{http://arxiv.org/abs/1206.5218}{{\tt arXiv:1206.5218 [hep-th]}}.
%%CITATION = ARXIV:1206.5218;%%.

\bibitem{ContributionDU}
T.~Dumitrescu, ``An Introduction to Supersymmetric Field Theories in Curved
  Space,'' {\em Journal of Physics A} {\bf xx} (2016)  000,
  \href{http://arxiv.org/abs/1608.02957}{{\tt 1608.02957}}.

\bibitem{PUFreedman:2013ryh}
D.~Z. Freedman and S.~S. Pufu, ``{The holography of $F$-maximization},''
  \href{http://dx.doi.org/10.1007/JHEP03(2014)135}{{\em JHEP} {\bf 03} (2014)
  135},
\href{http://arxiv.org/abs/1302.7310}{{\tt arXiv:1302.7310 [hep-th]}}.
%%CITATION = ARXIV:1302.7310;%%.

\bibitem{PUBorokhov:2002cg}
V.~Borokhov, A.~Kapustin, and X.-k. Wu, ``{Monopole operators and mirror
  symmetry in three-dimensions},''
  \href{http://dx.doi.org/10.1088/1126-6708/2002/12/044}{{\em JHEP} {\bf 12}
  (2002)  044},
\href{http://arxiv.org/abs/hep-th/0207074}{{\tt arXiv:hep-th/0207074
  [hep-th]}}.
%%CITATION = HEP-TH/0207074;%%.

\bibitem{PUBorokhov:2002ib}
V.~Borokhov, A.~Kapustin, and X.-k. Wu, ``{Topological disorder operators in
  three-dimensional conformal field theory},''
  \href{http://dx.doi.org/10.1088/1126-6708/2002/11/049}{{\em JHEP} {\bf 11}
  (2002)  049},
\href{http://arxiv.org/abs/hep-th/0206054}{{\tt arXiv:hep-th/0206054
  [hep-th]}}.
%%CITATION = HEP-TH/0206054;%%.

\bibitem{PUBenini:2009qs}
F.~Benini, C.~Closset, and S.~Cremonesi, ``{Chiral flavors and M2-branes at
  toric CY4 singularities},''
  \href{http://dx.doi.org/10.1007/JHEP02(2010)036}{{\em JHEP} {\bf 02} (2010)
  036},
\href{http://arxiv.org/abs/0911.4127}{{\tt arXiv:0911.4127 [hep-th]}}.
%%CITATION = ARXIV:0911.4127;%%.

\bibitem{PUJafferis:2009th}
D.~L. Jafferis, ``{Quantum corrections to $\mathcal{N} = 2$ Chern-Simons
  theories with flavor and their AdS$_{4}$ duals},''
  \href{http://dx.doi.org/10.1007/JHEP08(2013)046}{{\em JHEP} {\bf 08} (2013)
  046},
\href{http://arxiv.org/abs/0911.4324}{{\tt arXiv:0911.4324 [hep-th]}}.
%%CITATION = ARXIV:0911.4324;%%.

\bibitem{PUChester:2015qca}
S.~M. Chester, S.~Giombi, L.~V. Iliesiu, I.~R. Klebanov, S.~S. Pufu, and
  R.~Yacoby, ``{Accidental Symmetries and the Conformal Bootstrap},''
  \href{http://dx.doi.org/10.1007/JHEP01(2016)110}{{\em JHEP} {\bf 01} (2016)
  110},
\href{http://arxiv.org/abs/1507.04424}{{\tt arXiv:1507.04424 [hep-th]}}.
%%CITATION = ARXIV:1507.04424;%%.

\bibitem{PUChester:2015lej}
S.~M. Chester, L.~V. Iliesiu, S.~S. Pufu, and R.~Yacoby, ``{Bootstrapping
  $O(N)$ Vector Models with Four Supercharges in $3 \leq d \leq4$},''
\href{http://arxiv.org/abs/1511.07552}{{\tt arXiv:1511.07552 [hep-th]}}.
%%CITATION = ARXIV:1511.07552;%%.

\bibitem{PUChester:2014fya}
S.~M. Chester, J.~Lee, S.~S. Pufu, and R.~Yacoby, ``{The $ \mathcal{N}=8 $
  superconformal bootstrap in three dimensions},''
  \href{http://dx.doi.org/10.1007/JHEP09(2014)143}{{\em JHEP} {\bf 09} (2014)
  143},
\href{http://arxiv.org/abs/1406.4814}{{\tt arXiv:1406.4814 [hep-th]}}.
%%CITATION = ARXIV:1406.4814;%%.

\bibitem{PUBobev:2015jxa}
N.~Bobev, S.~El-Showk, D.~Mazac, and M.~F. Paulos, ``{Bootstrapping SCFTs with
  Four Supercharges},'' \href{http://dx.doi.org/10.1007/JHEP08(2015)142}{{\em
  JHEP} {\bf 08} (2015)  142},
\href{http://arxiv.org/abs/1503.02081}{{\tt arXiv:1503.02081 [hep-th]}}.
%%CITATION = ARXIV:1503.02081;%%.

\bibitem{PUBeem:2016cbd}
C.~Beem, W.~Peelaers, and L.~Rastelli, ``{Deformation quantization and
  superconformal symmetry in three dimensions},''
\href{http://arxiv.org/abs/1601.05378}{{\tt arXiv:1601.05378 [hep-th]}}.
%%CITATION = ARXIV:1601.05378;%%.

\bibitem{PUWess:1974tw}
J.~Wess and B.~Zumino, ``{Supergauge Transformations in Four-Dimensions},''
\href{http://dx.doi.org/10.1016/0550-3213(74)90355-1}{{\em Nucl. Phys.} {\bf
  B70} (1974)  39--50}.
%%CITATION = NUPHA,B70,39;%%.

\bibitem{PUStrassler:2003qg}
M.~J. Strassler, ``{An Unorthodox introduction to supersymmetric gauge
  theory},'' in {\em {Strings, Branes and Extra Dimensions: TASI 2001:
  Proceedings}}, pp.~561--638.
\newblock 2003.
\newblock
\href{http://arxiv.org/abs/hep-th/0309149}{{\tt arXiv:hep-th/0309149
  [hep-th]}}.
\newblock
%%CITATION = HEP-TH/0309149;%%.

\bibitem{PUFerreira:1997he}
P.~M. Ferreira and J.~A. Gracey, ``{The Beta function of the Wess-Zumino model
  at $O (1 / N^2)$},''
  \href{http://dx.doi.org/10.1016/S0550-3213(98)00236-3}{{\em Nucl. Phys.} {\bf
  B525} (1998)  435--456},
\href{http://arxiv.org/abs/hep-th/9712138}{{\tt arXiv:hep-th/9712138
  [hep-th]}}.
%%CITATION = HEP-TH/9712138;%%.

\bibitem{PUFerreira:1997hx}
P.~M. Ferreira and J.~A. Gracey, ``{Nonzeta knots in the renormalization of the
  Wess-Zumino model?},''
  \href{http://dx.doi.org/10.1016/S0370-2693(98)00169-5}{{\em Phys. Lett.} {\bf
  B424} (1998)  85--92},
\href{http://arxiv.org/abs/hep-th/9712140}{{\tt arXiv:hep-th/9712140
  [hep-th]}}.
%%CITATION = HEP-TH/9712140;%%.

\bibitem{PUFerreira:1996az}
P.~M. Ferreira, I.~Jack, and D.~R.~T. Jones, ``{The Quasiinfrared fixed point
  at higher loops},''
  \href{http://dx.doi.org/10.1016/S0370-2693(96)01549-3}{{\em Phys. Lett.} {\bf
  B392} (1997)  376--382},
\href{http://arxiv.org/abs/hep-ph/9610296}{{\tt arXiv:hep-ph/9610296
  [hep-ph]}}.
%%CITATION = HEP-PH/9610296;%%.

\bibitem{PUJack:1999fa}
I.~Jack and D.~R.~T. Jones, ``{Quasiinfrared fixed points and renormalization
  group invariant trajectories for nonholomorphic soft supersymmetry
  breaking},'' \href{http://dx.doi.org/10.1103/PhysRevD.61.095002}{{\em Phys.
  Rev.} {\bf D61} (2000)  095002},
\href{http://arxiv.org/abs/hep-ph/9909570}{{\tt arXiv:hep-ph/9909570
  [hep-ph]}}.
%%CITATION = HEP-PH/9909570;%%.

\bibitem{PUJack:1998iy}
I.~Jack, D.~R.~T. Jones, and A.~Pickering, ``{The soft scalar mass beta
  function},'' \href{http://dx.doi.org/10.1016/S0370-2693(98)00647-9}{{\em
  Phys. Lett.} {\bf B432} (1998)  114--119},
\href{http://arxiv.org/abs/hep-ph/9803405}{{\tt arXiv:hep-ph/9803405
  [hep-ph]}}.
%%CITATION = HEP-PH/9803405;%%.

\bibitem{PUnishioka2013rg}
T.~Nishioka and K.~Yonekura, ``On RG flow of $\tau_{RR}$ for supersymmetric
  field theories in three-dimensions,'' {\em Journal of High Energy Physics}
  {\bf 2013} (2013) no.~5, 1--20.

\bibitem{PUAharony:1997bx}
O.~Aharony, A.~Hanany, K.~A. Intriligator, N.~Seiberg, and M.~J. Strassler,
  ``{Aspects of ${\cal N}=2$ supersymmetric gauge theories in
  three-dimensions},''
  \href{http://dx.doi.org/10.1016/S0550-3213(97)00323-4}{{\em Nucl. Phys.} {\bf
  B499} (1997)  67--99},
\href{http://arxiv.org/abs/hep-th/9703110}{{\tt arXiv:hep-th/9703110
  [hep-th]}}.
%%CITATION = HEP-TH/9703110;%%.

\bibitem{PUWillett:2011gp}
B.~Willett and I.~Yaakov, ``{${\cal N}=2$ Dualities and $Z$ Extremization in
  Three Dimensions},''
\href{http://arxiv.org/abs/1104.0487}{{\tt arXiv:1104.0487 [hep-th]}}.
%%CITATION = ARXIV:1104.0487;%%.

\bibitem{PUKapustin:2010xq}
A.~Kapustin, B.~Willett, and I.~Yaakov, ``{Nonperturbative Tests of
  Three-Dimensional Dualities},''
  \href{http://dx.doi.org/10.1007/JHEP10(2010)013}{{\em JHEP} {\bf 10} (2010)
  013},
\href{http://arxiv.org/abs/1003.5694}{{\tt arXiv:1003.5694 [hep-th]}}.
%%CITATION = ARXIV:1003.5694;%%.

\bibitem{PUKapustin:2010mh}
A.~Kapustin, B.~Willett, and I.~Yaakov, ``{Tests of Seiberg-like Duality in
  Three Dimensions},''
\href{http://arxiv.org/abs/1012.4021}{{\tt arXiv:1012.4021 [hep-th]}}.
%%CITATION = ARXIV:1012.4021;%%.

\bibitem{PUAmariti:2011uw}
A.~Amariti, C.~Klare, and M.~Siani, ``{The Large $N$ Limit of Toric
  Chern-Simons Matter Theories and Their Duals},''
  \href{http://dx.doi.org/10.1007/JHEP10(2012)019}{{\em JHEP} {\bf 10} (2012)
  019},
\href{http://arxiv.org/abs/1111.1723}{{\tt arXiv:1111.1723 [hep-th]}}.
%%CITATION = ARXIV:1111.1723;%%.

\bibitem{PUSafdi:2012re}
B.~R. Safdi, I.~R. Klebanov, and J.~Lee, ``{A Crack in the Conformal Window},''
  \href{http://dx.doi.org/10.1007/JHEP04(2013)165}{{\em JHEP} {\bf 04} (2013)
  165},
\href{http://arxiv.org/abs/1212.4502}{{\tt arXiv:1212.4502 [hep-th]}}.
%%CITATION = ARXIV:1212.4502;%%.

\bibitem{PUKapustin:2011gh}
A.~Kapustin, ``{Seiberg-like duality in three dimensions for orthogonal gauge
  groups},''
\href{http://arxiv.org/abs/1104.0466}{{\tt arXiv:1104.0466 [hep-th]}}.
%%CITATION = ARXIV:1104.0466;%%.

\bibitem{PUJafferis:2011ns}
D.~Jafferis and X.~Yin, ``{A Duality Appetizer},''
\href{http://arxiv.org/abs/1103.5700}{{\tt arXiv:1103.5700 [hep-th]}}.
%%CITATION = ARXIV:1103.5700;%%.

\bibitem{PUKapustin:2011vz}
A.~Kapustin, H.~Kim, and J.~Park, ``{Dualities for 3d Theories with Tensor
  Matter},'' \href{http://dx.doi.org/10.1007/JHEP12(2011)087}{{\em JHEP} {\bf
  12} (2011)  087},
\href{http://arxiv.org/abs/1110.2547}{{\tt arXiv:1110.2547 [hep-th]}}.
%%CITATION = ARXIV:1110.2547;%%.

\bibitem{PUMorita:2011cs}
T.~Morita and V.~Niarchos, ``{F-theorem, duality and SUSY breaking in
  one-adjoint Chern-Simons-Matter theories},''
  \href{http://dx.doi.org/10.1016/j.nuclphysb.2012.01.003}{{\em Nucl. Phys.}
  {\bf B858} (2012)  84--116},
\href{http://arxiv.org/abs/1108.4963}{{\tt arXiv:1108.4963 [hep-th]}}.
%%CITATION = ARXIV:1108.4963;%%.

\bibitem{PUBenini:2011mf}
F.~Benini, C.~Closset, and S.~Cremonesi, ``{Comments on 3d Seiberg-like
  dualities},'' \href{http://dx.doi.org/10.1007/JHEP10(2011)075}{{\em JHEP}
  {\bf 10} (2011)  075},
\href{http://arxiv.org/abs/1108.5373}{{\tt arXiv:1108.5373 [hep-th]}}.
%%CITATION = ARXIV:1108.5373;%%.

\bibitem{PUGang:2011jj}
D.~Gang, C.~Hwang, S.~Kim, and J.~Park, ``{Tests of AdS$_4$/CFT$_3$
  correspondence for $\mathcal{N}=2$ chiral-like theory},''
  \href{http://dx.doi.org/10.1007/JHEP02(2012)079}{{\em JHEP} {\bf 02} (2012)
  079},
\href{http://arxiv.org/abs/1111.4529}{{\tt arXiv:1111.4529 [hep-th]}}.
%%CITATION = ARXIV:1111.4529;%%.

\bibitem{PUGulotta:2011aa}
D.~R. Gulotta, C.~P. Herzog, and S.~S. Pufu, ``{Operator Counting and
  Eigenvalue Distributions for 3D Supersymmetric Gauge Theories},''
  \href{http://dx.doi.org/10.1007/JHEP11(2011)149}{{\em JHEP} {\bf 11} (2011)
  149},
\href{http://arxiv.org/abs/1106.5484}{{\tt arXiv:1106.5484 [hep-th]}}.
%%CITATION = ARXIV:1106.5484;%%.

\bibitem{PUKim:2012vza}
H.~Kim and N.~Kim, ``{Operator Counting for ${\cal N}=2$ Chern-Simons Gauge
  Theories with Chiral-like Matter Fields},''
  \href{http://dx.doi.org/10.1007/JHEP05(2012)152}{{\em JHEP} {\bf 05} (2012)
  152},
\href{http://arxiv.org/abs/1202.6637}{{\tt arXiv:1202.6637 [hep-th]}}.
%%CITATION = ARXIV:1202.6637;%%.

\bibitem{PUMartelli:2011fu}
D.~Martelli, A.~Passias, and J.~Sparks, ``{The gravity dual of supersymmetric
  gauge theories on a squashed three-sphere},''
  \href{http://dx.doi.org/10.1016/j.nuclphysb.2012.07.019}{{\em Nucl. Phys.}
  {\bf B864} (2012)  840--868},
\href{http://arxiv.org/abs/1110.6400}{{\tt arXiv:1110.6400 [hep-th]}}.
%%CITATION = ARXIV:1110.6400;%%.

\bibitem{PUMartelli:2011fw}
D.~Martelli and J.~Sparks, ``{The gravity dual of supersymmetric gauge theories
  on a biaxially squashed three-sphere},''
  \href{http://dx.doi.org/10.1016/j.nuclphysb.2012.08.015}{{\em Nucl. Phys.}
  {\bf B866} (2013)  72--85},
\href{http://arxiv.org/abs/1111.6930}{{\tt arXiv:1111.6930 [hep-th]}}.
%%CITATION = ARXIV:1111.6930;%%.

\bibitem{PUMartelli:2012sz}
D.~Martelli, A.~Passias, and J.~Sparks, ``{The supersymmetric NUTs and bolts of
  holography},'' \href{http://dx.doi.org/10.1016/j.nuclphysb.2013.04.026}{{\em
  Nucl. Phys.} {\bf B876} (2013)  810--870},
\href{http://arxiv.org/abs/1212.4618}{{\tt arXiv:1212.4618 [hep-th]}}.
%%CITATION = ARXIV:1212.4618;%%.

\bibitem{PUMartelli:2013aqa}
D.~Martelli and A.~Passias, ``{The gravity dual of supersymmetric gauge
  theories on a two-parameter deformed three-sphere},''
  \href{http://dx.doi.org/10.1016/j.nuclphysb.2013.09.012}{{\em Nucl. Phys.}
  {\bf B877} (2013)  51--72},
\href{http://arxiv.org/abs/1306.3893}{{\tt arXiv:1306.3893 [hep-th]}}.
%%CITATION = ARXIV:1306.3893;%%.

\bibitem{PUFarquet:2014kma}
D.~Farquet, J.~Lorenzen, D.~Martelli, and J.~Sparks, ``{Gravity duals of
  supersymmetric gauge theories on three-manifolds},''
  \href{http://dx.doi.org/10.1007/JHEP08(2016)080}{{\em JHEP} {\bf 08} (2016)
  080},
\href{http://arxiv.org/abs/1404.0268}{{\tt arXiv:1404.0268 [hep-th]}}.
%%CITATION = ARXIV:1404.0268;%%.

\bibitem{PUFarquet:2013cwa}
D.~Farquet and J.~Sparks, ``{Wilson loops and the geometry of matrix models in
  AdS$_4$/CFT$_3$},'' \href{http://dx.doi.org/10.1007/JHEP01(2014)083}{{\em
  JHEP} {\bf 01} (2014)  083},
\href{http://arxiv.org/abs/1304.0784}{{\tt arXiv:1304.0784 [hep-th]}}.
%%CITATION = ARXIV:1304.0784;%%.

\bibitem{PUFarquet:2014bda}
D.~Farquet and J.~Sparks, ``{Wilson loops on three-manifolds and their M2-brane
  duals},'' \href{http://dx.doi.org/10.1007/JHEP12(2014)173}{{\em JHEP} {\bf
  12} (2014)  173},
\href{http://arxiv.org/abs/1406.2493}{{\tt arXiv:1406.2493 [hep-th]}}.
%%CITATION = ARXIV:1406.2493;%%.

\bibitem{PUBobev:2013cja}
N.~Bobev, H.~Elvang, D.~Z. Freedman, and S.~S. Pufu, ``{Holography for ${\cal
  N} = 2^*$ on $S^4$},'' \href{http://dx.doi.org/10.1007/JHEP07(2014)001}{{\em
  JHEP} {\bf 07} (2014)  001},
\href{http://arxiv.org/abs/1311.1508}{{\tt arXiv:1311.1508 [hep-th]}}.
%%CITATION = ARXIV:1311.1508;%%.

\bibitem{PUBobev:2016nua}
N.~Bobev, H.~Elvang, U.~Kol, T.~Olson, and S.~S. Pufu, ``{Holography for
  $\mathcal{N}=1^*$ on $S^4$},''
\href{http://arxiv.org/abs/1605.00656}{{\tt arXiv:1605.00656 [hep-th]}}.
%%CITATION = ARXIV:1605.00656;%%.

\bibitem{PUKlebanov:2012va}
I.~R. Klebanov, T.~Nishioka, S.~S. Pufu, and B.~R. Safdi, ``{Is Renormalized
  Entanglement Entropy Stationary at RG Fixed Points?},''
  \href{http://dx.doi.org/10.1007/JHEP10(2012)058}{{\em JHEP} {\bf 10} (2012)
  058},
\href{http://arxiv.org/abs/1207.3360}{{\tt arXiv:1207.3360 [hep-th]}}.
%%CITATION = ARXIV:1207.3360;%%.

\bibitem{PULiu:2013una}
H.~Liu and M.~Mezei, ``{Probing renormalization group flows using entanglement
  entropy},'' \href{http://dx.doi.org/10.1007/JHEP01(2014)098}{{\em JHEP} {\bf
  01} (2014)  098},
\href{http://arxiv.org/abs/1309.6935}{{\tt arXiv:1309.6935 [hep-th]}}.
%%CITATION = ARXIV:1309.6935;%%.

\bibitem{PUJafferis:2012iv}
D.~L. Jafferis and S.~S. Pufu, ``{Exact results for five-dimensional
  superconformal field theories with gravity duals},''
  \href{http://dx.doi.org/10.1007/JHEP05(2014)032}{{\em JHEP} {\bf 05} (2014)
  032},
\href{http://arxiv.org/abs/1207.4359}{{\tt arXiv:1207.4359 [hep-th]}}.
%%CITATION = ARXIV:1207.4359;%%.

\bibitem{PUFei:2014yja}
L.~Fei, S.~Giombi, and I.~R. Klebanov, ``{Critical $O(N)$ models in
  $6-\epsilon$ dimensions},''
  \href{http://dx.doi.org/10.1103/PhysRevD.90.025018}{{\em Phys. Rev.} {\bf
  D90} (2014) no.~2, 025018},
\href{http://arxiv.org/abs/1404.1094}{{\tt arXiv:1404.1094 [hep-th]}}.
%%CITATION = ARXIV:1404.1094;%%.

\end{thebibliography}
